\documentclass[aps, prc, floatfix, nofootinbib, superscriptaddress, twocolumn]{revtex4-1}

\usepackage{latexsym}
\usepackage{amsmath}
\usepackage{amssymb}
\usepackage{amsfonts}

\usepackage[mathscr,scaled=1.15]{urwchancal}
\DeclareFontFamily{OT1}{pzc}{}
\DeclareFontShape{OT1}{pzc}{m}{it}%
{<-> s * [1.15] pzcmi7t}{}
\DeclareMathAlphabet{\mathpzc}{OT1}{pzc}{m}{it}

\usepackage{color}

\usepackage{supertabular}
\usepackage{placeins}
\usepackage{epsfig}
\usepackage{graphicx}

\definecolor{purple}{rgb}{0.5,0,0.5}
\definecolor{blue}{rgb}{0.0,0,0.9}
\definecolor{prdblue}{rgb}{0.133,0.118,0.498}
\usepackage[colorlinks=true, pdfstartview=FitV, linkcolor=prdblue, citecolor= prdblue, urlcolor=prdblue]{hyperref}

\begin{document}


\title{Structure of the nucleon's low-lying excitations}



\author{Chen Chen}
\email[]{chenchen@ift.unesp.br}
\affiliation{Instituto de F\'isica Te\'orica, Universidade Estadual Paulista, Rua Dr.~Bento Teobaldo Ferraz, 271, 01140-070 S\~ao Paulo, SP, Brazil}

\author{Bruno~El-Bennich}
\email[]{bruno.bennich@cruzeirodosul.edu.br}
\affiliation{Universidade Cruzeiro do Sul, Rua Galv\~ao Bueno, 868, 01506-000 S\~ao Paulo, SP, Brazil}

\author{Craig D. Roberts}
\email[]{cdroberts@anl.gov}
\affiliation{Physics Division, Argonne National Laboratory, Argonne, Illinois
60439, USA}

\author{Sebastian M. Schmidt}
\email[]{s.schmidt@fz-juelich.de}
\affiliation{
Institute for Advanced Simulation, Forschungszentrum J\"ulich and JARA, D-52425 J\"ulich, Germany}

\author{\mbox{Jorge Segovia}}
\email[]{jsegovia@ifae.es}
\affiliation{
Institut de F\'{\i}sica d'Altes Energies (IFAE) and Barcelona Institute of Science and Technology (BIST),
Universitat Aut\`onoma de Barcelona, E-08193 Bellaterra (Barcelona), Spain
}

\author{Shaolong Wan}
\email[]{slwan@ustc.edu.cn}
\affiliation{Institute for Theoretical Physics and Department of Modern Physics,\\ University of Science and Technology of China, Hefei, Anhui 230026, P. R. China}

\date{08 November 2017}

\begin{abstract}
A continuum approach to the three valence-quark bound-state problem in quantum field theory is used to perform a comparative study of the four lightest $(I=1/2,J^P = 1/2^\pm)$ baryon isospin-doublets in order to elucidate their structural similarities and differences.  Such analyses predict the presence of nonpointlike, electromagnetically-active quark-quark (diquark) correlations within all baryons; and in these doublets, isoscalar-scalar, isovector-pseudovector, isoscalar-pseudoscalar, and vector diquarks can all play a role.
In the two lightest $(1/2,1/2^+)$ doublets, however, scalar and pseudovector diquarks are overwhelmingly dominant.  The associated rest-frame wave functions are largely $S$-wave in nature; and the first excited state in this $1/2^+$ channel has the appearance of a radial excitation of the ground state.
The two lightest $(1/2,1/2^-)$ doublets fit a different picture: accurate estimates of their masses are obtained by retaining only pseudovector diquarks; in their rest frames, the amplitudes describing their dressed-quark cores contain roughly equal fractions of even- and odd-parity diquarks; and the associated wave functions are predominantly $P$-wave in nature, but possess measurable $S$-wave components.  Moreover, the first excited state in each negative-parity channel has little of the appearance of a radial excitation.
In quantum field theory, all differences between positive- and negative-parity channels must owe to chiral symmetry breaking, which is overwhelmingly dynamical in the light-quark sector.  Consequently, experiments that can validate the contrasts drawn herein between the structure of the four lightest $(1/2,1/2^\pm)$ doublets will prove valuable in testing links between emergent mass generation and observable phenomena and, plausibly, thereby revealing dynamical features of confinement.
\end{abstract}



\maketitle


\section{Introduction}\label{introduction}
%
Developing a unified understanding of the four lightest $(I=1/2,J=1/2^\pm)$ baryon isospin-doublets in the hadron spectrum presents a challenging problem.  Whilst the proton is plainly a bound-state seeded by three valence-quarks, $u$, $u$, $d$, and the neutron is similar, the nature of the next three doublets: $N(1440)\,1/2^+$, $N(1535)\,1/2^-$, $N(1650)\,1/2^-$ is far less certain.  For example, the $N(1440)\,1/2^+$ ``Roper resonance'' \cite{Roper:1964zza, BAREYRE1964137, AUVIL196476, PhysRevLett.13.555, PhysRev.138.B190} has long been a source of puzzlement because a wide array of constituent-quark potential models produce a spectrum in which the second positive-parity doublet lies above the first negative-parity doublet.  This confusion was only resolved following \cite{Burkert:2017djo}: the acquisition and analysis of a vast amount of high-precision nucleon-resonance electroproduction data with single- and double-pion final states on a large kinematic domain of energy and photon virtuality, development of a sophisticated dynamical reaction theory capable of simultaneously describing all partial waves extracted from available, reliable data, and formulation and wide-ranging application of a Poincar\'e covariant approach to the continuum bound state problem in relativistic quantum field theory.  Today, it is widely judged that the Roper is, at heart, the first radial excitation of the nucleon, consisting of a well-defined dressed-quark core that is augmented by a meson cloud, which both reduces the Roper's core mass by approximately 20\% and contributes materially to the electroproduction transition form factors at low-$Q^2$.

%
Regarding the $N(1535)\,1/2^-$ and $N(1650)\,1/2^-$, an analogous picture ought to be correct.  However, new questions arise.  In constituent-quark models it is typical to describe these states as $P$-wave baryons \cite{Isgur:1978xj}, \emph{i.e}.\ quantum mechanical systems with one unit of constituent-quark orbital angular momentum, $L$, and classify them as members of the $(70,1_1^-)$ supermultiplet of $SU(3)\otimes O(3)$: the lighter state is associated with $L=1$, constituent-quark total spin $S=1/2$ coupled to $J=L+S=1/2$ and the heavier with $L=1$, $S=3/2$.  In relativistic quantum field theory, however, $L$ and $S$ are not good quantum numbers.  Moreover, even if they were, owing to the loss of particle number conservation, it is not clear \emph{a priori} just with which degrees-of-freedom $L$, $S$ should be connected.  This issue is related to the fact the constituent-quarks used in building quantum mechanical models have no known mathematical connection with the degrees-of-freedom featuring in quantum chromodynamics (QCD).  Plainly, there is still a great deal to learn about the nature of the nucleon's parity partner and its excitations.

The importance of this problem is all the greater because, in a symmetry-preserving treatment using relativistic quantum field theory, one may generate the interpolating field for the parity partner of any given state via a chiral rotation of that associated with the original state.  It follows that parity partners will be degenerate in mass and alike in structure in all theories that possess a chiral symmetry realised in the Wigner-Weyl mode.  (There is evidence of this, \emph{e.g}.\ in both continuum \cite{Maris:2000ig, Wang:2013wk} and lattice \cite{Cheng:2010fe, Aarts:2017rrl} analyses that explore the evolution of hadron properties with temperature.)
Such knowledge has long made the mass-splittings between parity partners in the strong-interaction spectrum a subject of interest.  The best known example, perhaps, is that provided by the $\rho(770)$- and $a_1(1260)$-mesons: viewed as chiral and hence parity partners, it has been argued \cite{Weinberg:1967kj} that their mass and structural differences can be attributed entirely to dynamical chiral symmetry breaking (DCSB), \emph{viz}.\ realisation of chiral symmetry in the Nambu-Goldstone mode.  It is plausible that this profound emergent feature of the Standard Model is tightly linked with confinement \cite{Roberts:2016vyn}; and regarding DCSB's role in explaining the splitting between parity partners, additional insights have been developed by studying the quantum field theory bound-state equations appropriate to the $\rho$- and $a_1$-mesons.  In their rest frames, one finds that their Poincar\'e-covariant wave functions are chiefly $S$-wave in nature \cite{Maris:1999nt, Chang:2008sp, Chang:2011ei, Roberts:2011cf, Chen:2012qr, Eichmann:2016yit}, even though both possess nonzero angular momentum \cite{Bloch:1999vka, Gao:2014bca}, whose magnitude influences the size of the splitting \cite{Chang:2011ei}.

Given the value of understanding the nature of the four lightest $(1/2,1/2^\pm)$ doublets in the hadron spectrum, herein we employ the methods of continuum quantum field theory in order to elucidate their structure.  One of our aims is to expose and clarify any commonalities that might exist between the quantum mechanical picture of these states and their character in QCD.  Complementing Ref.\,\cite{Segovia:2015hra}, the results will also prove useful in subsequent calculations of the $N(1535)\,1/2^-$ and $N(1650)\,1/2^-$ electroproduction form factors, existing empirical information on which \cite{Armstrong:1998wg, Thompson:2000by, Denizli:2007tq, Dalton:2008aa, Aznauryan:2009mx, Mokeev:2013kka} will be enlarged by forthcoming experiments at the Thomas Jefferson National Accelerator Facility (JLab).

We describe our approach to the baryon bound-state problem in Sec.\,\ref{secFaddeev}; detail and explain the character of the solutions for the four lightest $(1/2,1/2^\pm)$ doublets in Sec.\,\ref{solutions}; and summarise in Sec.\,\ref{epilogue}.

\section{Baryon Bound State Problem}
\label{secFaddeev}
\subsection{Faddeev equation}
\begin{figure}[t]
\centerline{%
\includegraphics[clip, width=0.45\textwidth]{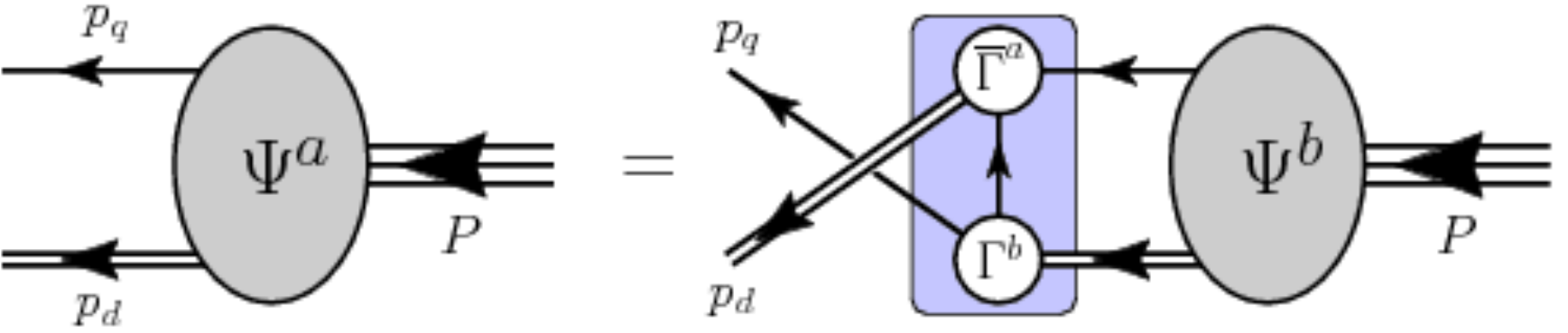}}
\caption{\label{figFaddeev}
Poincar\'e covariant Faddeev equation: a homogeneous linear integral equation for the matrix-valued function $\Psi$, being the Faddeev amplitude for a baryon of total momentum $P= p_q + p_d$, which expresses the relative momentum correlation between the dressed-quarks and -diquarks within the baryon.  The shaded rectangle demarcates the kernel of the Faddeev equation: \emph{single line}, dressed-quark propagator; $\Gamma$,  diquark correlation amplitude; and \emph{double line}, diquark propagator.}
\end{figure}

The problem of baryon structure in relativistic quantum field theory can be tackled using the Poincar\'e-covariant Faddeev equation introduced in Refs.\,\cite{Cahill:1988dx, Burden:1988dt, Cahill:1988zi, Reinhardt:1989rw, Efimov:1990uz}.  The Faddeev equation sums all possible exchanges and interactions that can take place between the three dressed-quarks that express the baryon's valence-quark content; and, used with a realistic quark-quark interaction \cite{Binosi:2014aea, Binosi:2016wcx, Binosi:2016nme}, it predicts the appearance of soft (nonpointlike) fully-interacting diquark correlations within baryons,
whose characteristics are greatly influenced by DCSB \cite{Segovia:2015ufa}.\footnote{Whilst the notion of diquark correlations was introduced long ago \cite{Lichtenberg:1967zz, Lichtenberg:1968zz}, the representation and understanding of these correlations has since evolved greatly, so that the dynamical correlations we exploit herein are vastly different from the pointlike constituents used in constituent spectroscopic models to analyse, \emph{e.g}.\ the missing resonance problem \cite{Santopinto:2016zkl}.}  Consequently, the problem of determining a baryon's internal structure is transformed into that of solving the linear, homogeneous matrix equation depicted in Fig.\,\ref{figFaddeev}.

\subsection{Dressed quarks}
In connection with the four lightest $(1/2,1/2^\pm)$ baryon doublets in the hadron spectrum, the kernel of the Faddeev equation in Fig.\,\ref{figFaddeev} involves three basic elements, \emph{viz}.\ the dressed light-quark propagator, $S(p)$, and the correlation amplitudes and propagators for all participating diquarks.
A great deal is known about $S(p)$, and in constructing the kernel we use the algebraic form described in Appendix~\ref{appendixKFE}, which has proven to be a very efficient parametrisation in the explanation and unification of a wide range of hadron observables \cite{Segovia:2014aza, Segovia:2015hra, Segovia:2016zyc}.
(\emph{N.B}.\ We assume isospin symmetry throughout, \emph{i.e}.\ the $u$- and $d$-quarks are mass-degenerate and described by the same propagator.  Moreover, all members of an isospin multiplet are also degenerate.)

\subsection{Correlation amplitudes}
Regarding the diquarks in Fig.\,\ref{figFaddeev}, all participating correlations are colour-antitriplets because they must combine with the bystander quark to form a colour singlet.  Notably, the colour-sextet quark+quark channel does not support correlations because gluon exchange is repulsive in this channel \cite{Cahill:1987qr}.

Diquark isospin-spin structure is more complex.  Accounting for Fermi-Dirac statistics, five types of correlation are possible in a $J=1/2$ bound-state: isoscalar-scalar ($I=0$, $J^P=0^+$), isovector-pseudo\-vector, isoscalar-pseudo\-scalar, iso\-scalar-vector, and iso\-vector-vector.  The leading structures in the correlation amplitudes for each case are, respectively:
{\allowdisplaybreaks
\begin{subequations}
\label{qqBSAs}
\begin{align}
\Gamma^{0^+}(k;K) & = g_{0^+} \, \gamma_5 C\, \tau^2 \,\vec{H} \,{\mathpzc F}(k^2/\omega_{0^+}^2) \,, \\
%
%
\vec{\Gamma}_\mu^{1^+}(k;K)
    & = i g_{1^+} \, \gamma_\mu C \, \vec{t}\, \vec{H} \,{\mathpzc F}(k^2/\omega_{1^+}^2)\,,\\
\Gamma^{0^-}(k;K) & = i g_{0^-} \, C\, \tau^2 \,\vec{H} \,{\mathpzc F}(k^2/\omega_{0^-}^2)\,,\\
%
%
\Gamma_\mu^{1^-}(k;K) & = g_{1^-} \, \gamma_\mu \gamma_5 C\, \tau^2 \,\vec{H} \,{\mathpzc F}(k^2/\omega_{1^-}^2)\,,\\
%
%
\vec{\Gamma}_\mu^{\bar{1}^-}(k;K)
    & = i g_{\bar{1}^-} \, [\gamma_\mu, \gamma\cdot K] \gamma_5 C \, \vec{t}\, \vec{H} \,{\mathpzc F}(k^2/\omega_{\bar{1}^-}^2) \,,
\end{align}
\end{subequations}}
\hspace*{-0.5\parindent}where:
$K$ is the total momentum of the correlation, $k$ is a two-body relative momentum, ${\mathpzc F}$ is the function in Eq.\,\eqref{defcalF}, $\omega_{J^P}$ is a size parameter, and $g_{J^P}$ is a coupling into the channel, which is fixed by normalisation;
%
%
$C=\gamma_2\gamma_4$ is the charge-conjugation matrix;
\begin{equation}
\label{tjmatrices}
\{t^j, j=+,0,-\} = \tfrac{1}{\surd 2} \{(\tau^0+\tau^3), \surd 2 \, \tau^1, (\tau^0-\tau^3)\} \,,
\end{equation}
$\tau^0=\,$diag$[1,1]$, $\{\tau^i,i=1,2,3\}$ are the Pauli matrices;
and $\vec{H} = \{i\lambda_c^7, -i\lambda_c^5,i\lambda_c^2\}$, with $\{\lambda_c^k,k=1,\ldots,8\}$ denoting Gell-Mann matrices in colour space, expresses the diquarks' colour antitriplet character.

The amplitudes in Eqs.\,\eqref{qqBSAs} are normalised canonically:
\begin{subequations}
\label{CanNorm}
\begin{align}
2 K_\mu & = \left. \frac{\partial }{\partial Q_\mu} \,  \Pi(K;Q)\right|_{Q=K}^{K^2 = - m_{J^P}^2},\\
\nonumber
\Pi(K;Q) & = {\rm tr} \int \frac{d^4 k}{(2\pi)^4} \bar \Gamma(k;-K) S(k+Q/2) \\
& \quad \times \Gamma(k;K) S^{\rm T}(-k+Q/2)\,, \label{eqPi}
\end{align}
\end{subequations}
where $\bar\Gamma(k;K) = C^\dagger \Gamma(-k;K) C$ and $[\cdot]^{\rm T}$ denotes matrix transpose.  When the correlation amplitudes involved carry Lorentz indices $\mu$, $\nu$, the left-hand-side of Eq.\,\eqref{eqPi} also includes a factor $\delta_{\mu\nu}$.  It is apparent now that the strength of coupling in each channel, $g_{J^P}$ in Eq.\,\eqref{qqBSAs}, is fixed by the associated value of $\omega_{J^P}$.

\subsection{Diquark propagators, masses, couplings}
A propagator is associated with each quark-quark correlation in Fig.\,\ref{figFaddeev}; and we use \cite{Segovia:2014aza}:
\begin{subequations}
\label{Eqqqprop}
\begin{align}
\Delta^{0^\pm}(K) & = \frac{1}{m_{0^\pm}^2} \, {\mathpzc F}(k^2/\omega_{0^\pm}^2)\,,\\
\Delta^{1^\pm}_{\mu\nu}(K) & = \left[ \delta_{\mu\nu} + \frac{K_\mu K_\nu}{m_{1^\pm}^2} \right]
 \frac{1}{m_{1^\pm}^2} \, {\mathpzc F}(k^2/\omega_{1^\pm}^2)\,.
\end{align}
\end{subequations}
These algebraic forms ensure that the diquarks are confined within the baryons, as appropriate for coloured correlations: whilst the propagators are free-particle-like at spacelike momenta, they are pole-free on the timelike axis; and this is sufficient to ensure confinement via the violation of reflection positivity (see, e.g.\ Ref.\,\cite{Horn:2016rip}, Sec.\,3).

The diquark masses and widths are related via \cite{Segovia:2014aza}
\begin{equation}
m_{J^P}^2 = 2 \,\omega_{J^P}^2,
\end{equation}
an identification which accentuates the free-particle-like propagation characteristics of the diquarks within the baryon.  The mass-scales are constrained by numerous studies; and we use (in GeV):
\begin{equation}
\label{dqmasses}
m_{0^+} = 0.8\,,\;
m_{1^+} = 0.9\,,\;
m_{0^-} = 1.2\,,\;
m_{1^-} = 1.3\,,
\end{equation}
where the first two values are drawn from Refs.\,\cite{Segovia:2014aza, Segovia:2015hra}, because they provide for a good description of numerous dynamical properties of the nucleon, $\Delta$-baryon and Roper resonance; and the masses of the odd-parity correlations are based on those computed in Ref.\,\cite{Lu:2017cln}.  (Such values are typical \cite{Eichmann:2016yit, Eichmann:2016hgl}; and in truncations of the two-body scattering problem that are most widely used, isoscalar- and isovector-vector correlations are degenerate.)  The impact of variations in these masses is readily estimated, \emph{e.g}.\ baryon masses typically respond linearly to changes in $m_{J^P}$ \cite{Roberts:2011cf}.

Using the values in Eq.\,\eqref{dqmasses} and Eqs.\,\eqref{qqBSAs}, \eqref{CanNorm}, one finds
\begin{equation}
\begin{array}{lll}
g_{0^+} = 14.8\,, &
g_{1^+} = 12.7\,,&
 \\
g_{0^-} = 12.8\,,&
g_{1^-} = 5.4\,,&
g_{{\bar 1}^-} = 2.5\,.
\end{array}
\end{equation}
Given that it is the coupling-squared which appears in the Faddeev kernels, then vector-diquark correlations should typically play a lesser role in the structure of $J=1/2$ baryons.  (There is some support for this expectation in Refs.\,\cite{Eichmann:2016jqx, Lu:2017cln}.)  Isovector-vector correlations should be especially suppressed because $g_{{\bar 1}^-}^2/g_{0^+}^2 = 0.03$.  Notably, too, isovector-vector diquarks are not supported in standard implementations of contact-interaction kernels \cite{Lu:2017cln}, which normally provide a good guide to the baryon spectrum.  Hereafter, therefore, we neglect isovector-vector correlations.  This expedient serves to simplify the Faddeev kernels: $22\times 22$ matrices are reduced to $16\times 16$.  (Here we also capitalise on isospin symmetry, which reduces the number of independent terms associated with isovector-pseudovector diquarks.)

\subsection{Remarks on the Faddeev kernels}
\label{FEremarks}
The elements described in the preceding subsections are sufficient to specify a Faddeev kernel in the $J^P=1/2^\pm$ channels associated with the four lightest $I=1/2$, $J=1/2$ baryon doublets.  Moreover, owing to our deliberate use of algebraic parametrisations for these inputs, the Faddeev equations thus obtained can be solved directly on the baryon mass-shells, providing simultaneously the associated on-shell Faddeev amplitudes and wave functions.

The inputs we use for the propagators and correlation amplitudes are constrained by observables and hence they express many effects that are lost in straightforward implementations of the lowest-order (rainbow-ladder, RL) truncation of the bound-state equations \cite{Binosi:2016rxz}.  Notwithstanding that, some correction of the Faddeev kernels is necessary to overcome an intrinsic weakness of the equation depicted in Fig.\,\ref{figFaddeev}, whose structure is based on the rainbow-ladder truncation.
Solving this equation, one finds that ground-state positive-parity octet baryons are primarily constituted from like-parity diquarks, with negligible contributions from negative-parity correlations.  This makes sense.  However, the parity partners of the ground-state baryons are also overwhelmingly dominated by the positive-parity diquarks and, consequently, too light.  It is possible that something important is missing.

A failure to generate adequate splitting between parity partners is a familiar flaw: the masses of odd-parity mesons are also too low when computed in RL truncation; and the cause there is an absence of spin-orbit repulsion owing to an oversimplification of the gluon-quark vertex when formulating the RL bound-state equations \cite{Chang:2009zb, Chang:2011ei, Binosi:2014aea, Binosi:2016rxz, Eichmann:2016yit}.  It is now possible to solve meson bound-state problems with more sophisticated kernels, which incorporate crucial nonperturbative mechanisms and hence better express the role of spin-orbit repulsion and related effects.  However, that is not yet the case in the baryon sector; and we therefore employ a simple artifice in order to implement the missing interactions.  Namely \cite{Lu:2017cln}, we introduce a single parameter into the Faddeev equation for $J^P=(1/2)^P$ baryons: $g_{\rm DB}$, a linear multiplicative factor attached to each opposite-parity $(-P)$ diquark amplitude in the baryon's Faddeev equation kernel.  For example, in the Faddeev kernel for the $N(1535)\,1/2^-$, each entry is multiplied by $g_{DB}^n$, where $n$ counts the number of positive-parity diquark correlation amplitudes that are present.  $g_{\rm DB}$ is the single free parameter in our study.

\subsection{Faddeev amplitudes and wave functions}
In solving the Faddeev equation, Fig.\,\ref{figFaddeev}, one obtains both the mass-squared and bound-state amplitude of all baryons with a given value of $J^P$.  In fact, it is the form of the Faddeev amplitude which fixes the channel.  A $1/2^\pm$ baryon is described by
\begin{align}
\Psi^{\pm} & = \psi^{\pm}_1 + \psi^{\pm}_2 + \psi^{\pm}_3\,,
\end{align}
where the subscript identifies the bystander quark, \emph{i.e}.\ the quark that is not participating in a diquark correlation,  $\psi^{\pm}_{1,2}$ are obtained from $\psi^{\pm}_3=:\psi^{\pm}$ by a cyclic permutation of all quark labels, and
\begin{align}
\nonumber
\psi^\pm&(p_i,\alpha_i,\sigma_i) \\
\nonumber
& = [\Gamma^{0^+}(k;K)]^{\alpha_1 \alpha_2}_{\sigma_1 \sigma_2} \, \Delta^{0^+}(K) \,[\varphi_{0^+}^\pm(\ell;P) u(P)]^{\alpha_3}_{\sigma_3} \\
\nonumber
& \quad +   [\Gamma^{1^+j}_\mu] \, \Delta_{\mu\nu}^{1^+}\, [\varphi_{1^+ \nu }^{j\pm}(\ell;P) u(P)] \\
\nonumber
& \quad+   [\Gamma^{0^- }]\, \Delta^{0^-}\, [\varphi_{0^-}^\pm(\ell;P) u(P)] \\
& \quad+   [\Gamma^{1^-}_\mu] \,\Delta_{\mu\nu}^{1^-}\, [\varphi_{1^- \nu}^\pm(\ell;P) u(P)] \,,
\label{FaddeevAmp}
\end{align}
where
$(p_i,\sigma_i,\alpha_i)$ are the momentum, spin and isospin labels of the quarks constituting the bound state;
$P=p_1 + p_2 + p_3=p_d+p_q$ is the total momentum of the baryon;
$k=(p_1-p_2)/2$, $K=p_1+p_2=p_d$, $\ell = (-K + 2 p_3)/3$; $j$ is the label in Eq.\,\eqref{tjmatrices}; and $u(P)$ is a Euclidean spinor (see Ref.\,\cite{Segovia:2014aza}, Appendix\,B for details).  The remaining elements in Eq.\,\eqref{FaddeevAmp} are the following matrix-valued functions:
\begin{subequations}
\label{sapv}
\begin{align}
\varphi_{0^+}^\pm(\ell;P) & = \sum_{i=1}^2 {\mathpzc s}_i^{\pm}(\ell^2,\ell\cdot P)\,  {\mathpzc S}^i(\ell;P) \,{\mathpzc G}^\pm\,, \\
\varphi_{1^+ \nu}^{j \pm}(\ell;P)  & = \sum_{i=1}^6 {\mathpzc a}_i^{j \pm}(\ell^2,\ell\cdot P)\, \gamma_5 {\mathpzc A}^i_\nu(\ell;P)\, {\mathpzc G}^\pm \,, \\
\varphi_{0^-}^\pm(\ell;P) &= \sum_{i=1}^2 {\mathpzc p}_i^{\pm}(\ell^2,\ell\cdot P)\,  {\mathpzc S}^i(\ell;P)\, {\mathpzc G}^\mp\,,\\
\varphi_{1^- \nu}^{\pm}(\ell;P)  & = \sum_{i=1}^6 {\mathpzc v}_i^{\pm}(\ell^2,\ell\cdot P)\, \gamma_5  {\mathpzc A}_\nu^i(\ell;P)\, {\mathpzc G}^\mp  \,,
\end{align}
\end{subequations}
where ${\mathpzc G}^{+(-)} = {\mathbf I}_{\rm D} \, (\gamma_5)$ and
\begin{align}
\nonumber
{\mathpzc S}^1 & = {\mathbf I}_{\rm D} \,,\;
{\mathpzc S}^2  = i \gamma\cdot\hat\ell - \hat\ell \cdot\hat P {\mathbf I}_{\rm D} \\
{\mathpzc A}^1_\nu & =  \gamma\cdot\ell^\perp \hat P_\nu\,,\;
{\mathpzc A}^2_\nu  = - i \hat P_\nu {\mathbf I}_{\rm D}\,,\;
{\mathpzc A}^3_\nu  = \gamma\cdot\hat\ell^\perp \hat\ell^\perp_\nu \\
\nonumber
{\mathpzc A}^4_\nu & = i\hat \ell_\nu^\perp {\mathbf I}_{\rm D}\,,\;
{\mathpzc A}^5_\nu = \gamma_\nu^\perp - {\mathpzc A}^3_\nu\,,\;
{\mathpzc A}^6_\nu = i \gamma_\nu^\perp \gamma\cdot\hat\ell^\perp - {\mathpzc A}^4_\nu\,,
\end{align}
with $\hat\ell^2=1$, $\hat P^2 = -1$, $\ell^\perp = \hat\ell_\nu +\hat\ell\cdot\hat P \hat P_\nu$, $\gamma^\perp = \gamma_\nu +\gamma\cdot\hat P \hat P_\nu$.

The (unamputated) Faddeev wave function can be computed from the amplitude specified by Eqs.\,\eqref{FaddeevAmp}, \eqref{sapv} simply by attaching the appropriate dressed-quark and diquark propagators.  It may also be decomposed in the form of Eqs.\,\eqref{sapv}.  Naturally, the scalar functions are different, and we label them $\tilde{\mathpzc s}_i^{\pm}$, $\tilde{\mathpzc a}_i^{j \pm}$, $\tilde {\mathpzc p}_i^{\pm}$, $\tilde{\mathpzc v}_i^{\pm}$.

Both the Faddeev amplitude and wave function are Poincar\'e covariant, \emph{i.e}.\ they are qualitatively identical in all reference frames.  Naturally, each of the scalar functions that appears is frame-independent, but the frame chosen determines just how the elements should be combined.  In consequence, the manner by which the dressed-quarks' spin, $S$, and orbital angular momentum, $L$, add to form $J=1/2^P$ is frame-dependent: $L$, $S$ are not independently Poincar\'e invariant.\footnote{The nature of the combination is also scale dependent because the definition of a dressed-quark and the character of the correlation amplitudes changes with resolving scale, $\zeta$, in a well-defined manner \cite{Lepage:1980fj}.  Our analysis is understood to be valid at $\zeta \simeq 1\,$GeV.}
Hence, in order to enable comparisons with typical formulations of constituent quark models, here we list the set of baryon rest-frame quark-diquark angular momentum identifications \cite{Oettel:1998bk, Cloet:2007piS}:
{\allowdisplaybreaks
\begin{subequations}
\label{Lidentifications}
\begin{align}
^2\!S: & \quad {\mathpzc S}^1, {\mathpzc A}^2_\nu, ({\mathpzc A}^3_\nu+{\mathpzc A}^5_\nu) \,,\\
%
^2\!P: & \quad {\mathpzc S}^2, {\mathpzc A}^1_\nu, ({\mathpzc A}^4_\nu+{\mathpzc A}^6_\nu)\,,\\
%
^4\!P: & \quad (2{\mathpzc A}^4_\nu-{\mathpzc A}^6_\nu)/3\,,\\
%
^4\!D:  & \quad  (2{\mathpzc A}^3_\nu-{\mathpzc A}^5_\nu)/3  \,,
\end{align}
\end{subequations}}
\hspace*{-0.5\parindent}\emph{viz}.\ the scalar functions associated with these combinations of Dirac matrices in a Faddeev wave function possess the identified angular momentum correlation between the quark and diquark.
Those functions are:
\begin{subequations}
\label{LFunctionIdentifications}
\begin{align}
\label{LFunctionIdentificationsa}
^2\!S: & \quad \tilde{\mathpzc s}_1^\pm, \tilde{\mathpzc a}_2^\pm,
    (\tilde{\mathpzc a}_3^\pm+2{\mathpzc a}_5^\pm)/3\,,\\
^2\!P: & \quad \tilde{\mathpzc s}_2^\pm, \tilde{\mathpzc a}_1^\pm,
    (\tilde{\mathpzc a}_4^\pm+2\tilde{\mathpzc a}_6^\pm)/3\,,\\
^4\!P: & \quad (\tilde{\mathpzc a}_4^\pm-\tilde{\mathpzc a}_6^\pm)\,,\\
^4\!D: & \quad (\tilde{\mathpzc a}_3^\pm - \tilde{\mathpzc a}_5^\pm) \,,
\end{align}
\end{subequations}
with analogous associations for $\{ \tilde{\mathpzc p}_i^\pm, i=1,2 \}$, $\{ \tilde{\mathpzc v}_i^\pm , i=1,\ldots,6\}$.

\section{Solutions and their Properties}
\label{solutions}
\subsection{Masses of the dressed-quark cores}
Using the information provided in Sec.\,\ref{secFaddeev}, it is straightforward to generalise the procedures detailed, \emph{e.g}.\ in Ref.\,\cite{Segovia:2014aza}, and obtain the linear, homogeneous matrix integral equations satisfied by the Faddeev amplitudes of $(1/2,1/2^\pm)$ baryons.  Capitalising on isospin symmetry, there are just two equations, corresponding to $P=\pm$, and herein we are interested in the two lowest-mass solutions of each equation.  (In the absence of chiral symmetry breaking, dynamical and explicit, these two equations are indistinguishable.)

We have one parameter, \emph{i.e}.\ $g_{\rm DB}$, described in Sec.\,\ref{FEremarks}; and we choose $g_{\rm DB}=0.43$ so as to produce a mass-splitting of $0.1\,$GeV between the lowest-mass $P=-$ state and the first excited $P=+$ state, \emph{viz}.\ the empirical value.

Our computed values for the masses of the four lightest $1/2^\pm$ baryon doublets are listed here, in GeV:
\begin{equation}
\label{eqMasses}
\begin{array}{l|cccc}
g_{\rm DB} & m_N & m_{N(1440)}^{1/2^+} & m_{N(1535)}^{1/2^-} & m_{N(1650)}^{1/2^-}\\\hline
0.43 & 1.19 & 1.73 & 1.83 & 1.91 \\
1.0  & 1.19 & 1.73 & 1.43 & 1.61 \\
0.0  & 1.19 & 1.73 & 2.16 & 2.31
\end{array}\,.
\end{equation}
In order to understand the results, recall that $g_{\rm DB}=0$ ensures $P=-(+)$ diquarks are eliminated from the Faddeev kernel of $P=+(-)$ baryons, whereas $g_{\rm DB}=1$ means they appear with unmodified strength.  Evidently, therefore, pseudoscalar and vector diquarks have no impact on the mass of the two positive-parity baryons, whereas scalar and pseudovector diquarks are important to the negative parity systems.  (A $\pm 10$\% change in $g_{\rm DB}$ around our fitted value only alters each $P=-$ mass by less than $2$\%.)
It is worth noting, too, that although $1/2^-$ solutions exist even if one eliminates isoscalar-pseudoscalar and -vector diquarks, $1/2^+$ solutions do not exist in the absence of scalar and pseudovector diquarks.  The first clause here is, perhaps, surprising. It indicates that, with our Faddeev kernel, the so-called $P$-wave (negative-parity) baryons can readily be built from positive-parity diquarks.  This indicates that the energy-cost associated with introducing quark-diquark orbital angular momentum is not very high.  As noted elsewhere \cite{Lu:2017cln}, in the absence of a true beyond-RL Faddeev kernel, one cannot judge whether this is a veracious feature of the strong interaction within baryons or an artefact of existing kernels.

It is now worth highlighting that the kernel in Fig.\,\ref{figFaddeev} omits all those resonant contributions which may be associated with meson-baryon final-state interactions that are resummed in dynamical coupled channels models  in order to transform a bare-baryon into the observed state \cite{JuliaDiaz:2007kz, Suzuki:2009nj, Kamano:2010ud, Ronchen:2012eg, Kamano:2013iva, Doring:2014qaa}.  The Faddeev equations analysed to produce the results in Eq.\,\eqref{eqMasses} should therefore be understood as producing the \emph{dressed-quark core} of the bound-state, not the completely-dressed and hence observable object \cite{Eichmann:2008ae, Eichmann:2008ef}.  In consequence, a comparison between the empirical values of the resonance pole positions and the masses in Eq.\,\eqref{eqMasses} is not pertinent.  Instead, one should compare the masses of the quark core with values determined for the meson-undressed bare-excitations, \emph{e.g}.:
\begin{equation}
\begin{array}{l|llll}
& m_N & m_{N(1440)}^{1/2^+} & m_{N(1535)}^{1/2^-} & m_{N(1650)}^{1/2^-}
\\\hline
{\rm herein} & 1.19 & 1.73 & 1.83 & 1.91
\\
M_{B}^0 \; \mbox{\cite{Suzuki:2009nj}}  & & 1.76 & 1.80 & 1.88 
\end{array},
\end{equation}
where $M_B^0$ is the relevant bare mass inferred in the associated dynamical coupled-channels analysis \cite{Suzuki:2009nj}.  Notably, the rms-relative-difference between our predicted quark core masses and the bare-masses determined in Ref.\,\cite{Suzuki:2009nj} is just 1.7\%, even though no attempt was made to secure agreement.  We consider this to be a success of our formulation of the bound-state problem for a baryon's dressed-quark core.

\subsection{Rest-frame orbital angular momentum}
It is interesting now to dissect the results in various ways and thereby sketch the character of the quark cores that constitute the four lightest $1/2^\pm$ baryon doublets.  We begin with an exposition of their rest-frame orbital angular momentum content, to which purpose we compiled Table~\ref{tableL}.
Plainly, the nucleon and $N(1440)\,1/2^+$ are primarily $S$-wave in nature, since they are not supported by the Faddeev equation unless $S$-wave components are contained in the wave function.  On the other hand, the $N_0^-=N(1535)\,1/2^-$, $N_1^-= N(1650)\,1/2^-$ are essentially $P$-wave in character.  These observations provide support in quantum field theory for the constituent-quark model classifications of these systems, so long as angular momentum is understood at the hadronic scale to be that between the quark and diquark.

\begin{table}[t]
\caption{\label{tableL}
Computed quark-core masses of the low-lying $1/2^\pm$ baryons.
Row\,1: results obtained using the complete Faddeev wave function, \emph{i.e}.\ with all angular momentum components included.
Subsequent rows: masses obtained when the indicated rest-frame angular momentum component(s) is(are) excluded from the Faddeev wave function.
Empty locations indicate that a solution is not obtained under the conditions indicated.
Legend: $N_0^+$ is the ground-state nucleon, $N_1^+ = N(1440)\,1/2^+$, $N_0^-= N(1535)\,1/2^-$, $N_1^-= N(1650)\,1/2^-$.
(All dimensioned quantities are listed in GeV.)}
\begin{center}
\begin{tabular*}
{\hsize}
{
c@{\extracolsep{0ptplus1fil}}
|c@{\extracolsep{0ptplus1fil}}
c@{\extracolsep{0ptplus1fil}}
c@{\extracolsep{0ptplus1fil}}
c@{\extracolsep{0ptplus1fil}}
|c@{\extracolsep{0ptplus1fil}}
c@{\extracolsep{0ptplus1fil}}}\hline
 & \multicolumn{4}{c|}{$g_{\rm DB}=0.43$} & \multicolumn{2}{c}{$g_{\rm DB}=1.0$} \\
 $L\,$content & $N_0^+$ & $N_1^+$ & $N_0^-$ & $N_1^-$ & $N_0^-$ & $N_1^-$ \\\hline
$S,P,D$ & 1.19 & 1.73 & 1.83 & 1.91\; & 1.43 & 1.61  \\\hline
$-,P,D$ &         &        & 1.89 & 1.98\; & 1.55 & 1.75  \\
$S,-, D$ & 1.24 & 1.71 &       &           &         &         \\
$S, P, -$ &1.20 & 1.74 & 1.83 & 1.91\; & 1.44 & 1.61 \\\hline
$S, -, -$ & 1.24 & 1.71 &        &           &        &         \\
$-, P, -$ &        &        & 1.90 & 1.98 \; & 1.57 & 1.75 \\\hline
\end{tabular*}
\end{center}
\end{table}

To elucidate, we turn our attention to the Faddeev wave functions themselves.  Connected with each matrix in Eqs.\,\eqref{Lidentifications}, there is a scalar function, the collection of which we denote as $\{{\mathpzc E}_i^\pm,i=1,\ldots,8\}$, \emph{e.g}.\ the six rest-frame $^2\!S$-components in a $P=+$ baryon are connected with ${\mathpzc E}_{1,2,3}^\pm$ and, using Eq.\,\eqref{LFunctionIdentificationsa}, these functions are
$\{
(\tilde{\mathpzc s}_1^+, \tilde{\mathpzc p}_1^+),
(\tilde{\mathpzc a}_2^+,\tilde{\mathpzc v}_2^+),
([\tilde{\mathpzc a}_3^+ +2\tilde{\mathpzc a}_5^+]/3, [\tilde{\mathpzc v}_3^++2\tilde{\mathpzc v}_5^+]/3) \}$.
For each baryon, we compute
\begin{equation}
{\mathpzc L}_i^\pm = \int \frac{d^4 \ell}{(2\pi)^4} \, |  {\mathpzc E}_{i}^\pm(\ell^2,\ell\cdot P)|^2
\end{equation}
and subsequently define the following rest-frame angular momentum strengths:
{\allowdisplaybreaks
\begin{subequations}
\label{Lfracs}
\begin{align}
{\mathbb S} & ={\mathbb T}^{-1} \, \sum_{k=\pm} \sum_{i \in \, ^2\!S} {\mathpzc L}_i^k \,,\\
{\mathbb P} & ={\mathbb T}^{-1} \, \sum_{k=\pm} \sum_{i \in \, ^2\!P,\, ^4\!P} {\mathpzc L}_i^k \,,\\
{\mathbb D} & ={\mathbb T}^{-1} \, \sum_{k=\pm} \sum_{i \in \,  ^4\!D} {\mathpzc L}_i^k \,,\\
{\mathbb T} & =\sum_{k=\pm} \sum_{i=1}^8\, {\mathpzc L}_i^k\,.
\end{align}
\end{subequations}
}

\begin{figure}[t]
\centerline{%
\includegraphics[clip, width=0.45\textwidth]{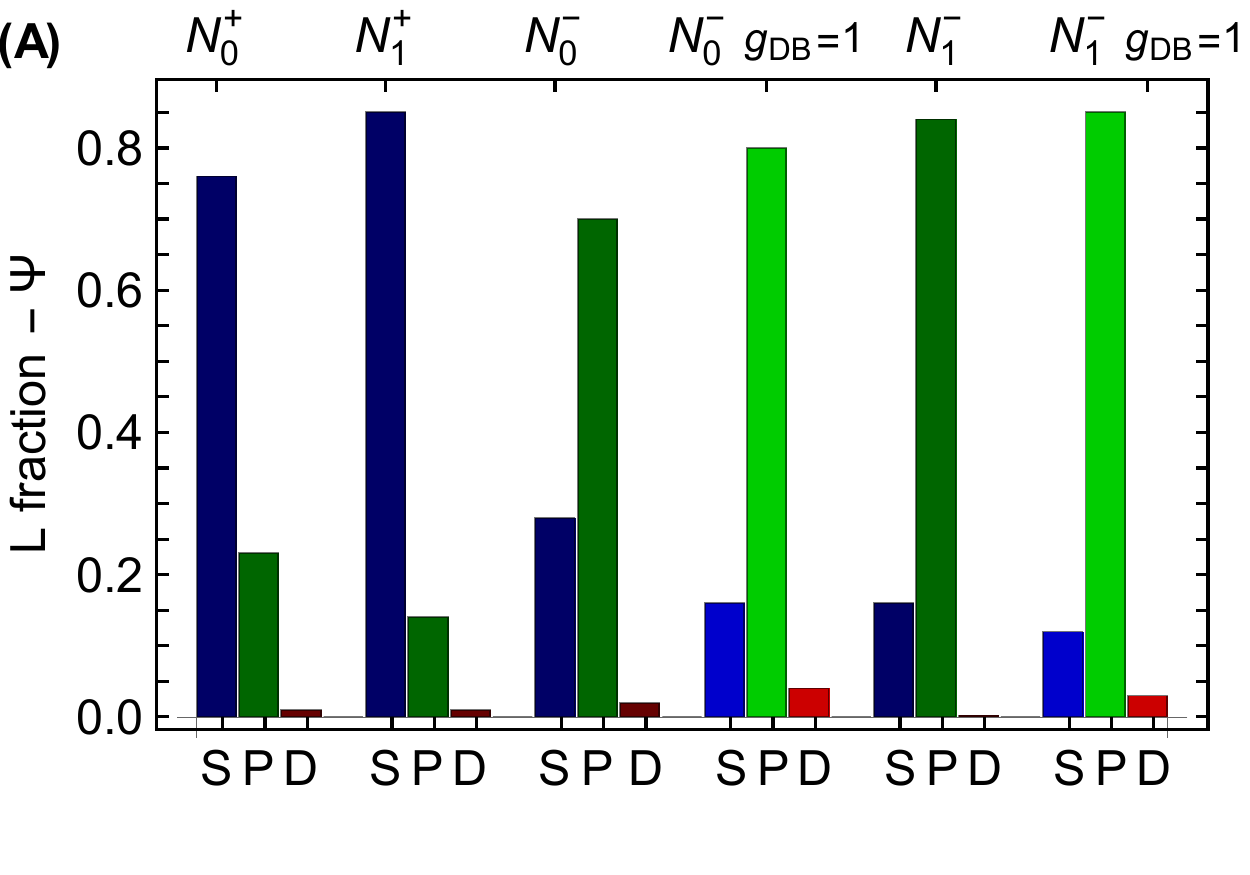}}
\centerline{%
\includegraphics[clip, width=0.45\textwidth]{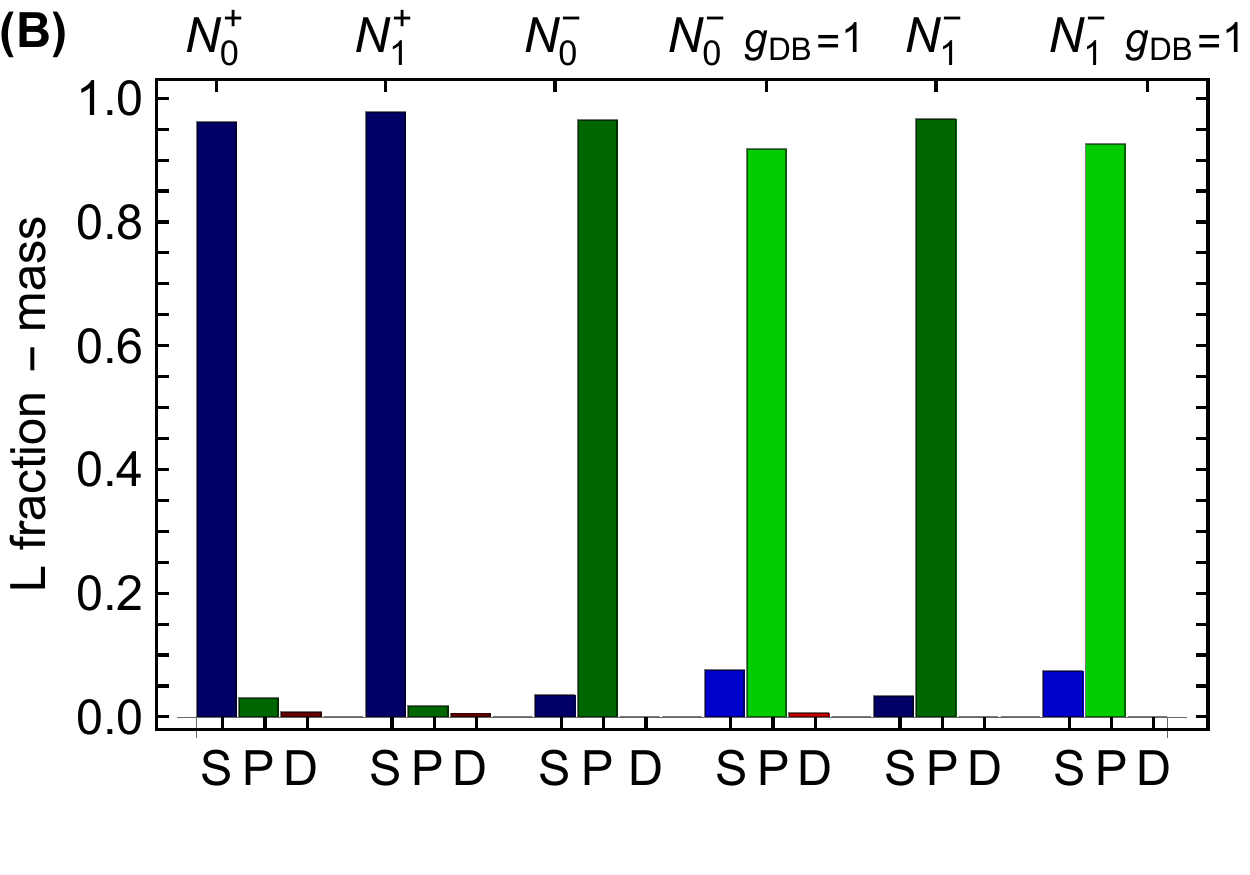}}
\caption{\label{BarLD}
\emph{Upper panel} -- (A) Baryon rest-frame quark-di\-quark orbital angular momentum fractions, as defined in Eqs.\,\eqref{Lfracs}.
\emph{Lower panel} -- (B) Relative contribution of various quark-diquark orbital angular momentum components to the mass of a given baryon. 
In both panels, the results were computed with $g_{\rm DB}=0.43$, except for the identified bar-triplets with lighter shading, for which $g_{\rm DB}=1$.
Legend: $N_0^+$ is the ground-state nucleon, $N_1^+ = N(1440)\,1/2^+$, $N_0^-= N(1535)\,1/2^-$, $N_1^-= N(1650)\,1/2^-$.
}
\end{figure}

Our computed values for the rest-frame quark-diquark angular momentum fractions are depicted in Fig.\,\ref{BarLD}A.  As telegraphed by Table~\ref{tableL}, in their rest frames, the two lightest $1/2^+$ doublets are predominantly $S$-wave in character, whereas the negative parity states are chiefly $P$-wave.  Evidently, $g_{\rm DB}<1$ has the effect of suppressing the $P$-wave component in the negative-parity baryons for reasons we will subsequently elucidate.
In all cases the $D$-wave components are negligible.

It is interesting to note that if one makes all diquarks equally massive, setting $m_{qq}=1.2\,$GeV, then $S$-waves are enhanced in $1/2^+$ systems, whereas the $D$-wave component becomes larger in $1/2^-$ systems at the cost of a roughly 10\% reduction in the sum of $S$- and $P$-waves.
Plainly, details of baryon internal structure are sensitive to the size and ordering of diquark masses.

Another, perhaps better, way to sketch the relative importance of different partial waves within a baryon is to depict their contributions to a given observable.  Herein, we consider the mass.  Using the information in Table\,\ref{tableL}, in Fig.\,\ref{BarLD}B we depict the relative contribution to a hadron's mass that derives from a given angular momentum component in the baryon's rest-frame Faddeev wave function.  For the purpose of this illustration, we draw all bars as positive, even though it is usually the case that the dominant partial wave produces a baryon with mass greater than the final result and adding an angular momentum component introduces attraction.  As an example, the nucleon entries are drawn from:
\begin{equation}
\{ S\! :  1.24, \; P\! : 1.24-1.20,\; D\! : 1.20-1.19 \}/\mathpzc{T}\,,
\end{equation}
$\mathpzc{T}=(1.24+0.04+0.01)\,$GeV.  This measure delivers the same qualitative picture of each baryon's internal structure as that presented in Fig.\,\ref{BarLD}A: apparently, therefore, there is little mixing between partial waves in the computation of a baryon's mass.

\subsection{Diquark content}
Accounting for isospin symmetry in the pseudovector diquark component, the Faddeev amplitude of each baryon is readily decomposed into a sum of sixteen distinct terms, $\{{\mathpzc F}_i,i=1,\ldots,16\}$, each one of which is uniquely identified with a particular diquark type.  In connection with each term, we define
\begin{equation}
{\mathpzc D}_i = \int\frac{d^4\ell}{(2\pi)^4}\,|{\mathpzc F}_i(\ell^2,\ell\cdot P)|^2
\end{equation}
and subsequently compute
\begin{equation}
\label{Dfracs}
{\mathbb Q}_{\mathpzc t}  ={\mathbb S}^{-1} \,  \sum_{i \in {\mathpzc t}} {\mathpzc D}_i \,,\;
{\mathbb S}  = \sum_{i=1}^{16}\, {\mathpzc D}_i\,,
%
%
%
%
%
\end{equation}
where ${\mathpzc t}={\mathpzc s}(=0^+)$, ${\mathpzc a}(=1^+)$, ${\mathpzc p}(=0^-)$, ${\mathpzc v}(=1^-)$.
The values of these rest-frame fractions are one indication of the relative strengths of the various diquark components within a baryon.

\begin{figure}[t]
\centerline{%
\includegraphics[clip, width=0.45\textwidth]{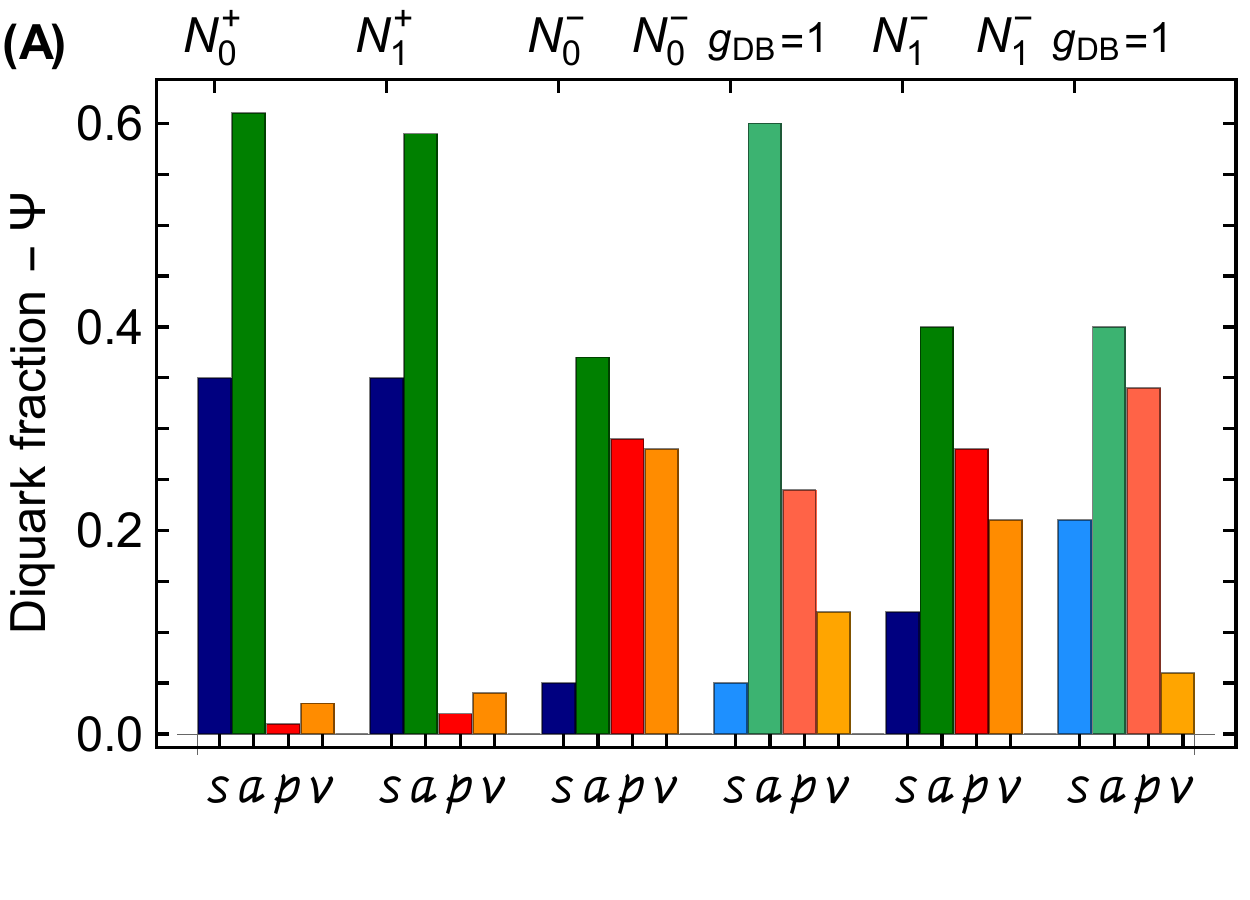}}
\centerline{%
\includegraphics[clip, width=0.45\textwidth]{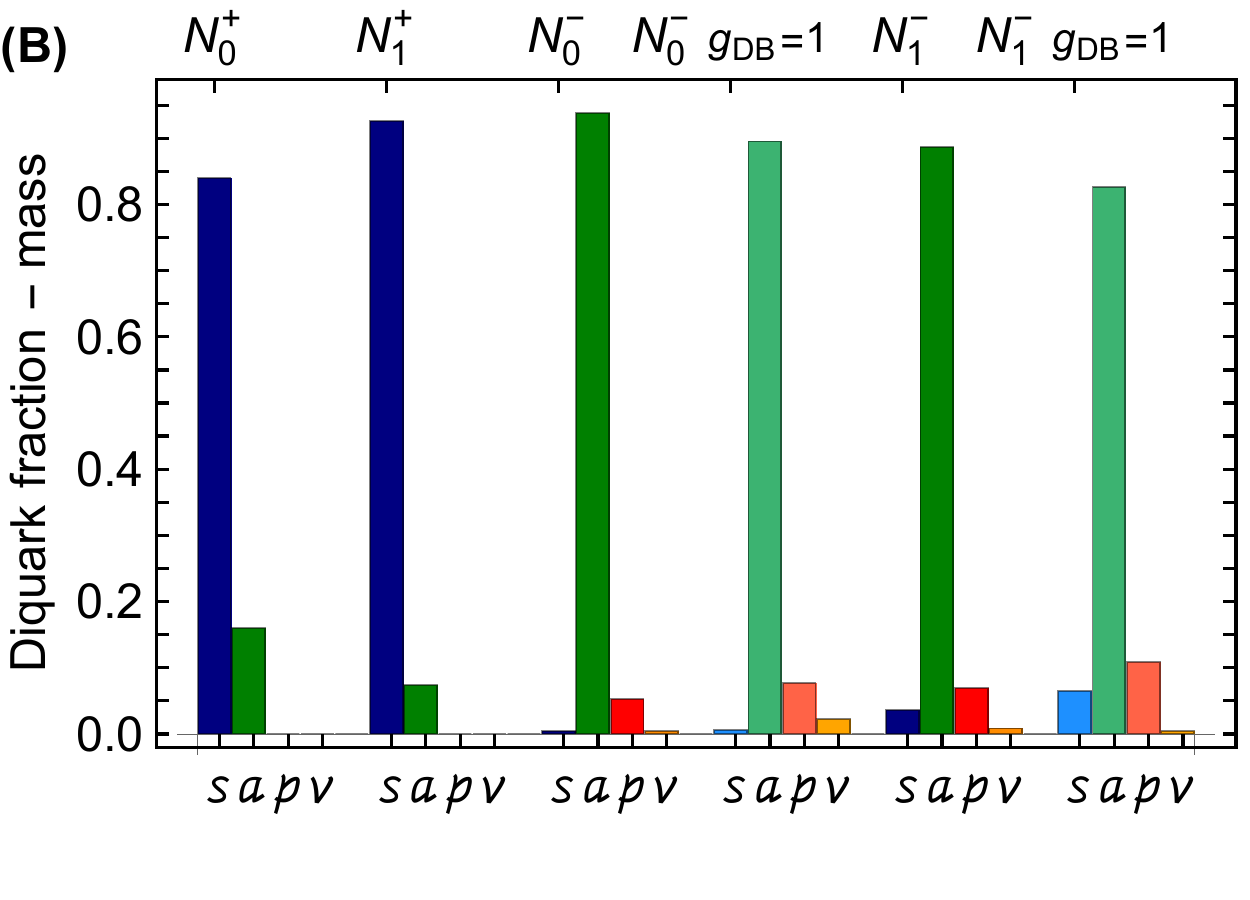}}
\caption{\label{BarLDN}
\emph{Upper panel} -- (A) Relative strengths of various diquark components within the indicated baryon's Faddeev amplitude, as defined in Eqs.\,\eqref{Dfracs}.
%
\emph{Lower panel} -- (B) Relative contribution to a baryon's mass from a given diquark correlation in that baryon's Faddeev amplitude.
In both panels, the results were computed with $g_{\rm DB}=0.43$, except for the identified bar-quadruplets with lighter shading, for which $g_{\rm DB}=1$.
Legend: $N_0^+$ is the ground-state nucleon, $N_1^+ = N(1440)\,1/2^+$, $N_0^-= N(1535)\,1/2^-$, $N_1^-= N(1650)\,1/2^-$.
}
\end{figure}

\begin{figure*}[!t]
\begin{center}
\begin{tabular}{lr}
\includegraphics[clip,width=0.4\linewidth]{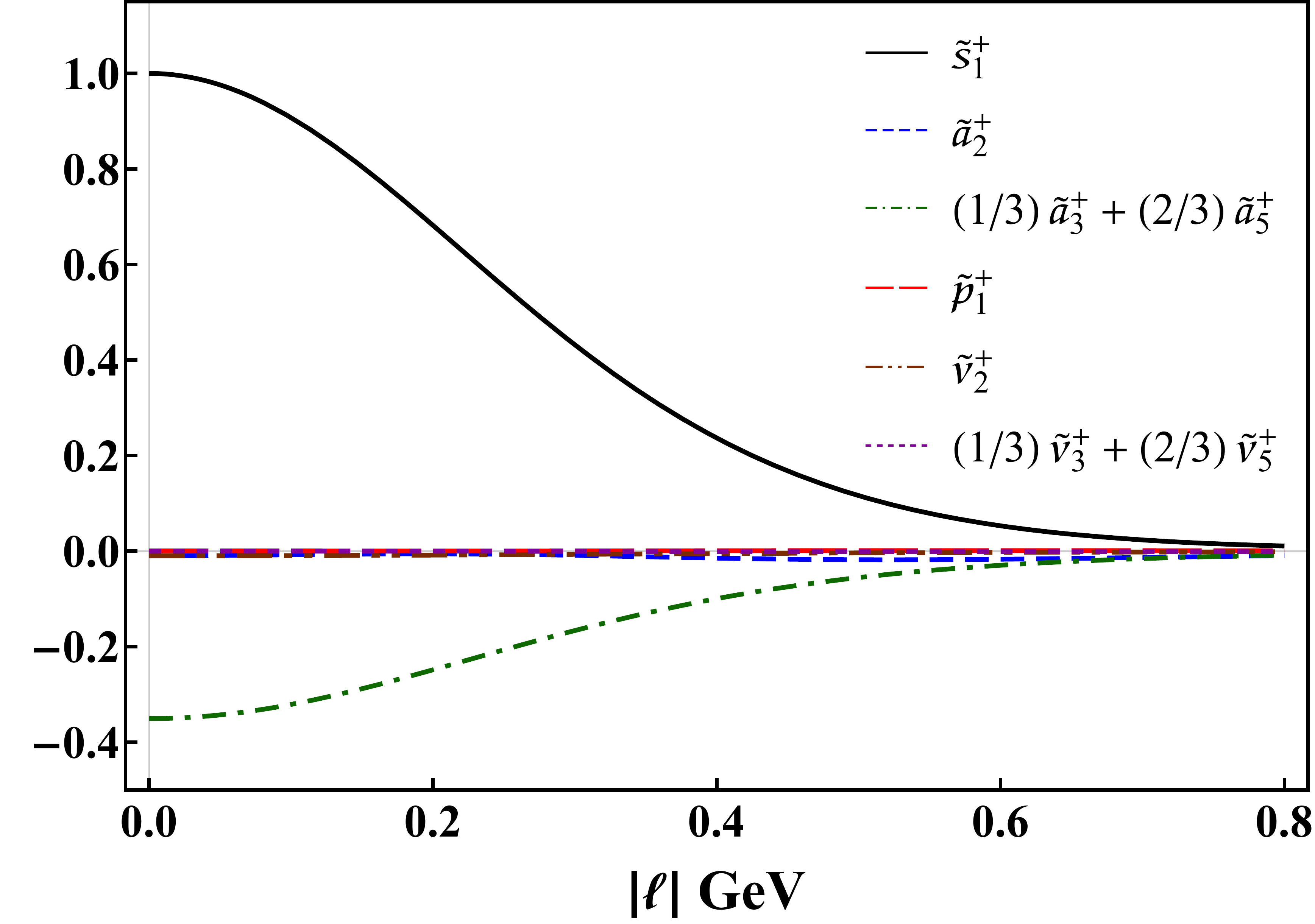}\hspace*{2ex } &
\includegraphics[clip,width=0.4\linewidth]{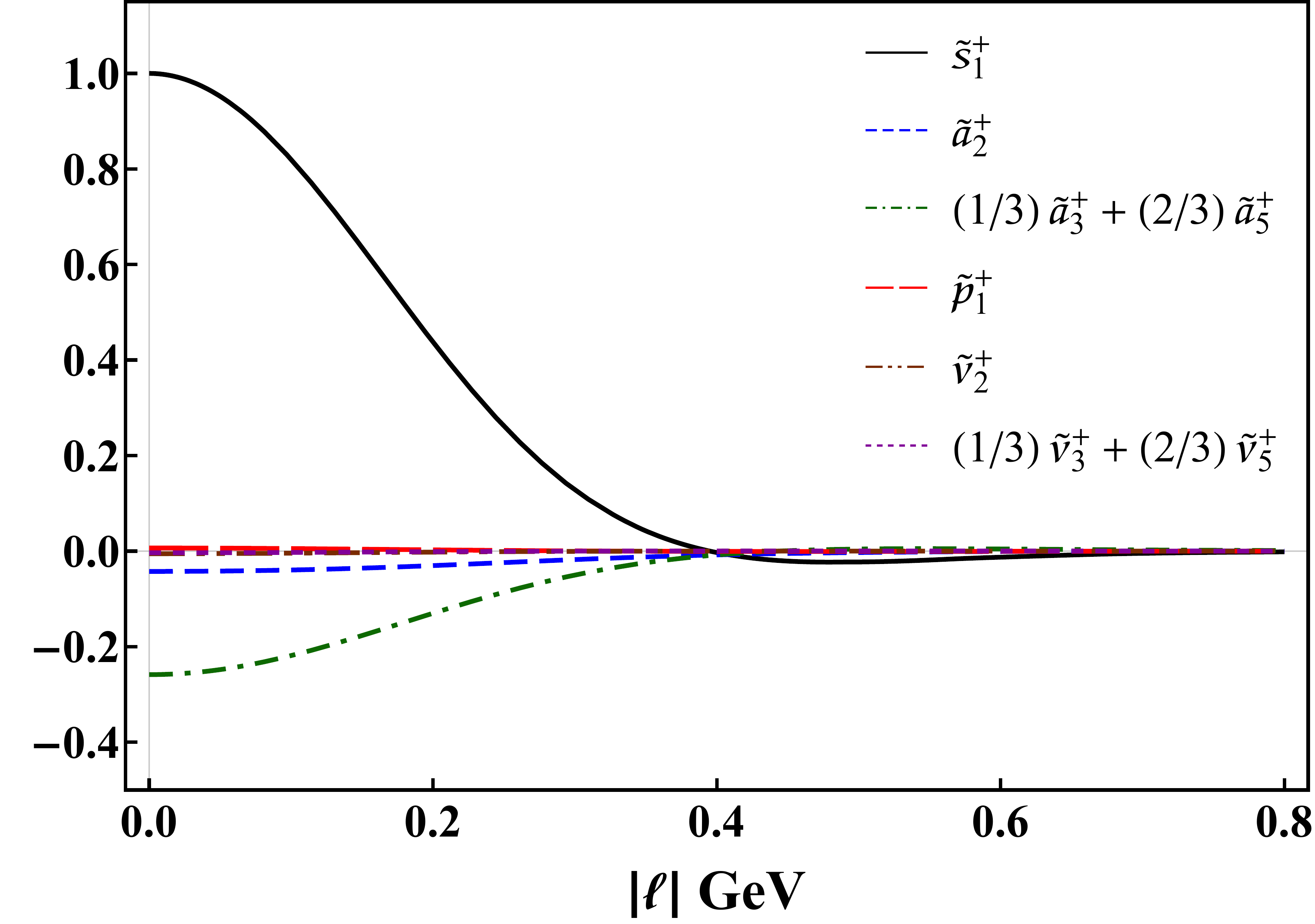}\vspace*{-0ex}
\end{tabular}
\begin{tabular}{lr}
\includegraphics[clip,width=0.4\linewidth]{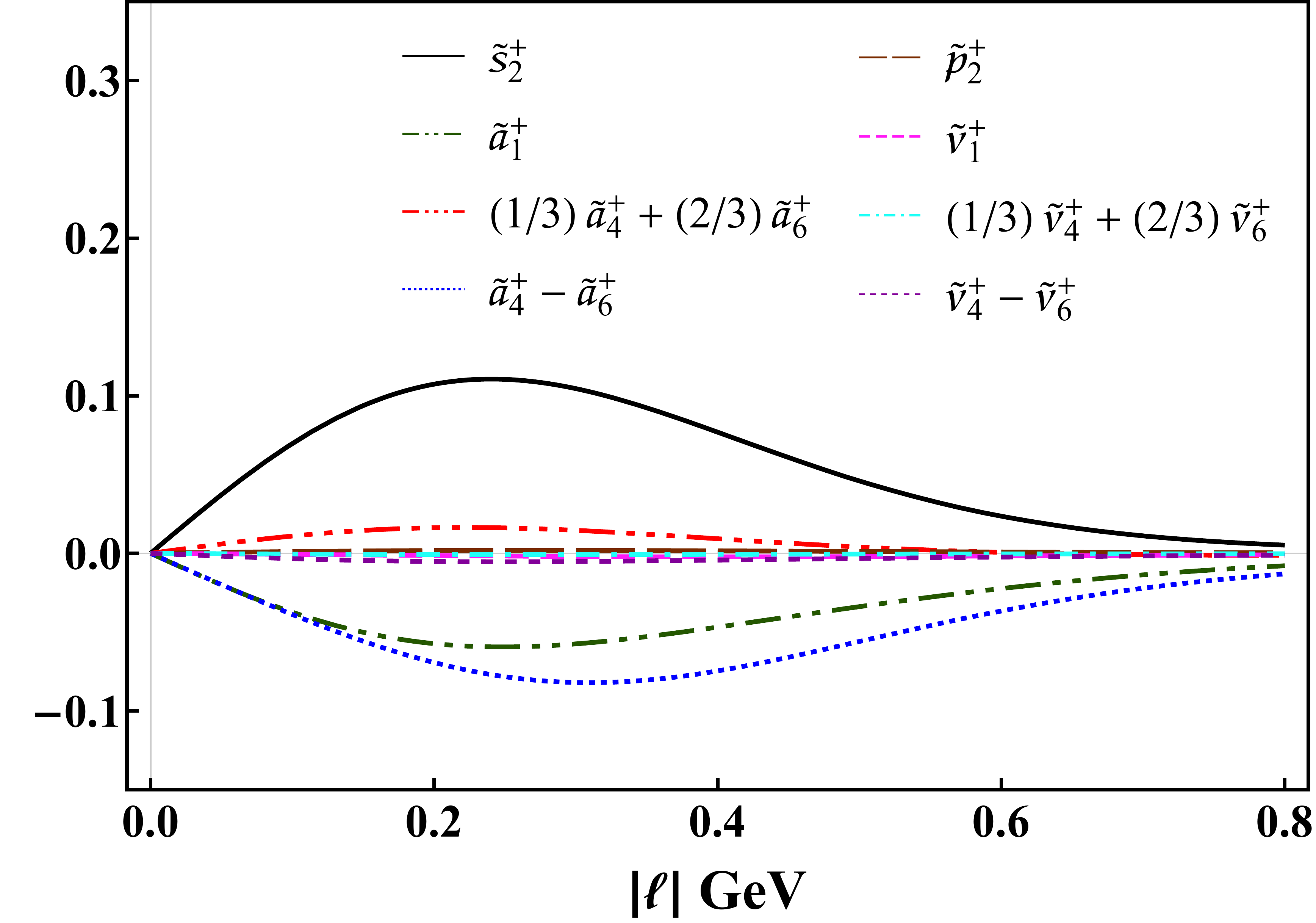}\hspace*{2ex } &
\includegraphics[clip,width=0.4\linewidth]{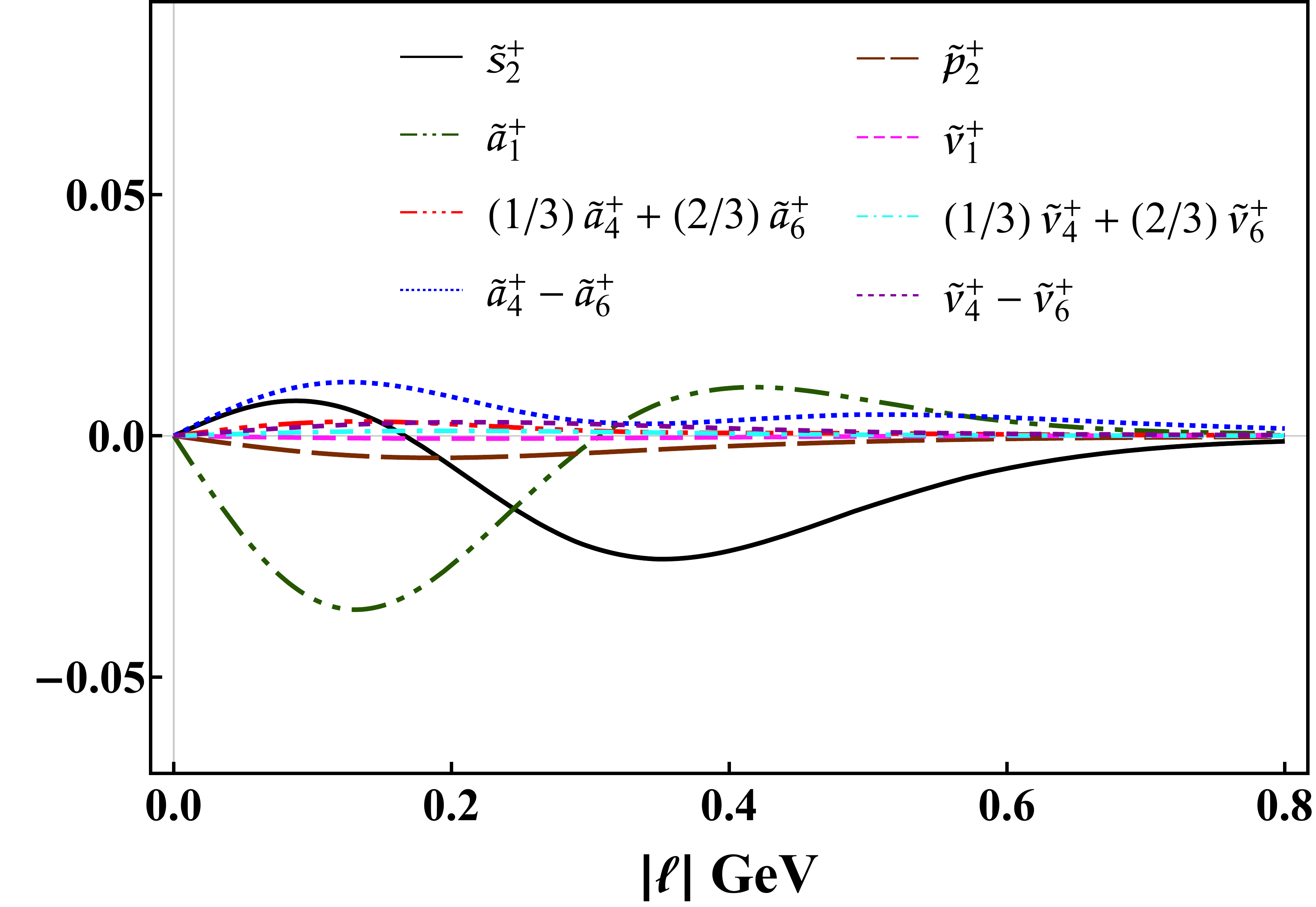}\vspace*{-1ex}
\end{tabular}
\end{center}
\caption{\label{FAPp}
Order-zero Chebyshev projections, Eq.\,\eqref{Wproject}, of the nucleon (left) and Roper (right) quark-core Faddeev wave functions, with $S$-wave in the top row and $P$-wave in the bottom.  For a given baryon, all functions are rescaled by the associated $\ell^2=0$ value of the zeroth moment of $\tilde {\mathpzc s}_1^+$. }
\end{figure*}

Our results for the diquark fractions defined by Eqs.\,\eqref{Dfracs} are depicted in Fig.\,\ref{BarLDN}A.
This measure indicates that $g_{\rm DB}<1$ has little impact on the nucleon and Roper, so we do not draw $g_{\rm DB}=1$ results.  On the other hand, it has a significant effect on the structure of the negative parity baryons, serving to enhance the net negative-parity diquark content.

One can now explain the impact of $g_{\rm DB}<1$ on the rest-frame quark-diquark angular momentum fractions in negative parity baryons.   Within such systems, it increases both the effective energy-cost (mass) of positive parity diquarks and the fraction of pseudoscalar- and vector-diquarks they contain.  Each of these effects serves individually to lower the total rest-frame angular momentum, and they are mutually reinforcing.

It is worth remarking that if one makes all diquarks equally massive, setting $m_{qq}=1.2\,$GeV, then the isoscalar-vector fraction is enhanced in $1/2^+$ baryons, at the cost of a $\sim 30$\% reduction in the sum of scalar and pseudovector fractions, whereas the pseudoscalar and vector diquark fractions in negative-parity baryons both increase substantially, so that they become dominant, at the cost of the same size reduction in the sum of scalar and pseudovector fractions.  These changes highlight once again that details of baryon structure are sensitive to the size and ordering of diquark masses.

Diquark fractions for the nucleon and Roper resonance, computed using the same Faddeev equations, are presented in Ref.\,\cite{Segovia:2015hra}.  The measure used therein is different, based on the Faddeev amplitudes' canonical normalisation, which is a Poincar\'e invariant quantity related to baryon number.  There are similarities, \emph{e.g}.\ using either scheme, the nucleon and Roper possess very similar diquark content; and differences, \emph{e.g}.\ using the normalisation measure, the scalar diquark is dominant.  The latter emphasises that in the computation of an observable quantity, there is significant interference between the distinct diquark components in a baryon's Faddeev amplitude.  One learns from these observations that comparisons between diquark fractions computed for different baryons using the same indicator are easily interpreted, whereas that is not always the case for comparisons between results obtained for the same baryon using different schemes.

In order to draw a closer connection herein with the standard used in Ref.\,\cite{Segovia:2015hra}, in Fig.\,\ref{BarLDN} we depict the relative contributions to a hadron's mass owing to each of the diquark components in the baryon's Faddeev amplitude.  Here, the difference between the upper and lower panels of Fig.\,\ref{BarLDN} is marked.  In each case depicted in the lower panel, there is a single dominant diquark component; and each new correlation adds binding, reducing the computed mass.  In some cases, \emph{e.g}.\ the subleading ${\mathpzc s}$ and ${\mathpzc p}$ correlations in $N_1^-$, there is significant constructive interference.   Measuring the relative strength of diquark correlations through their contribution to a baryon's mass and the canonical normalisation, one arrives at an understanding which is quite different from that suggested by Fig.\,\ref{BarLDN}A, \emph{viz}.\ to a fair degree of accuracy, a range of observable nucleon and Roper properties are largely determined by their scalar diquark content and those of the lightest states in the negative-parity channel are primarily fixed by their pseudovector diquark content.

\begin{figure*}[!ht]
\begin{center}
\begin{tabular}{lr}
\includegraphics[clip,width=0.4\linewidth]{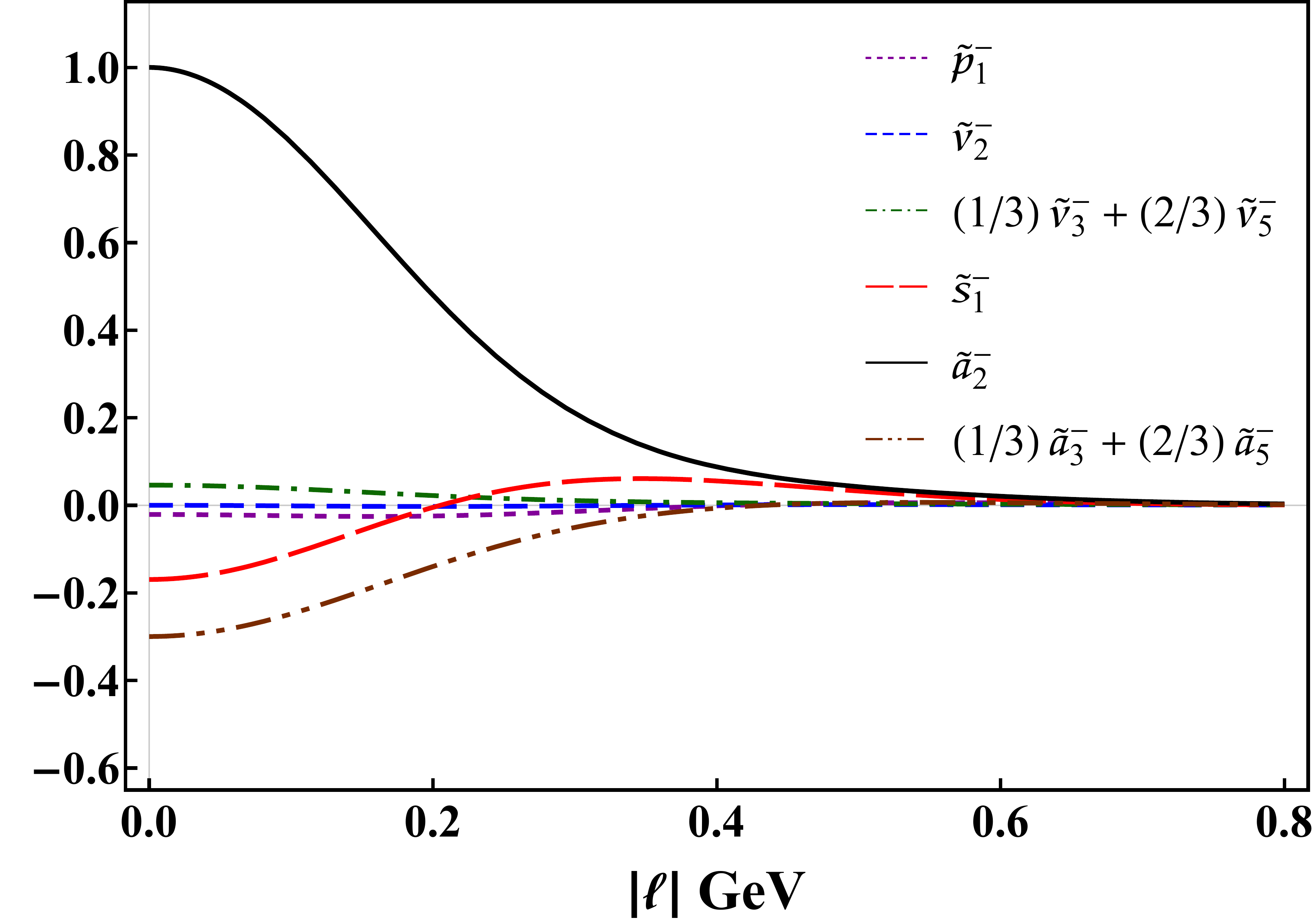}\hspace*{2ex } &
\includegraphics[clip,width=0.4\linewidth]{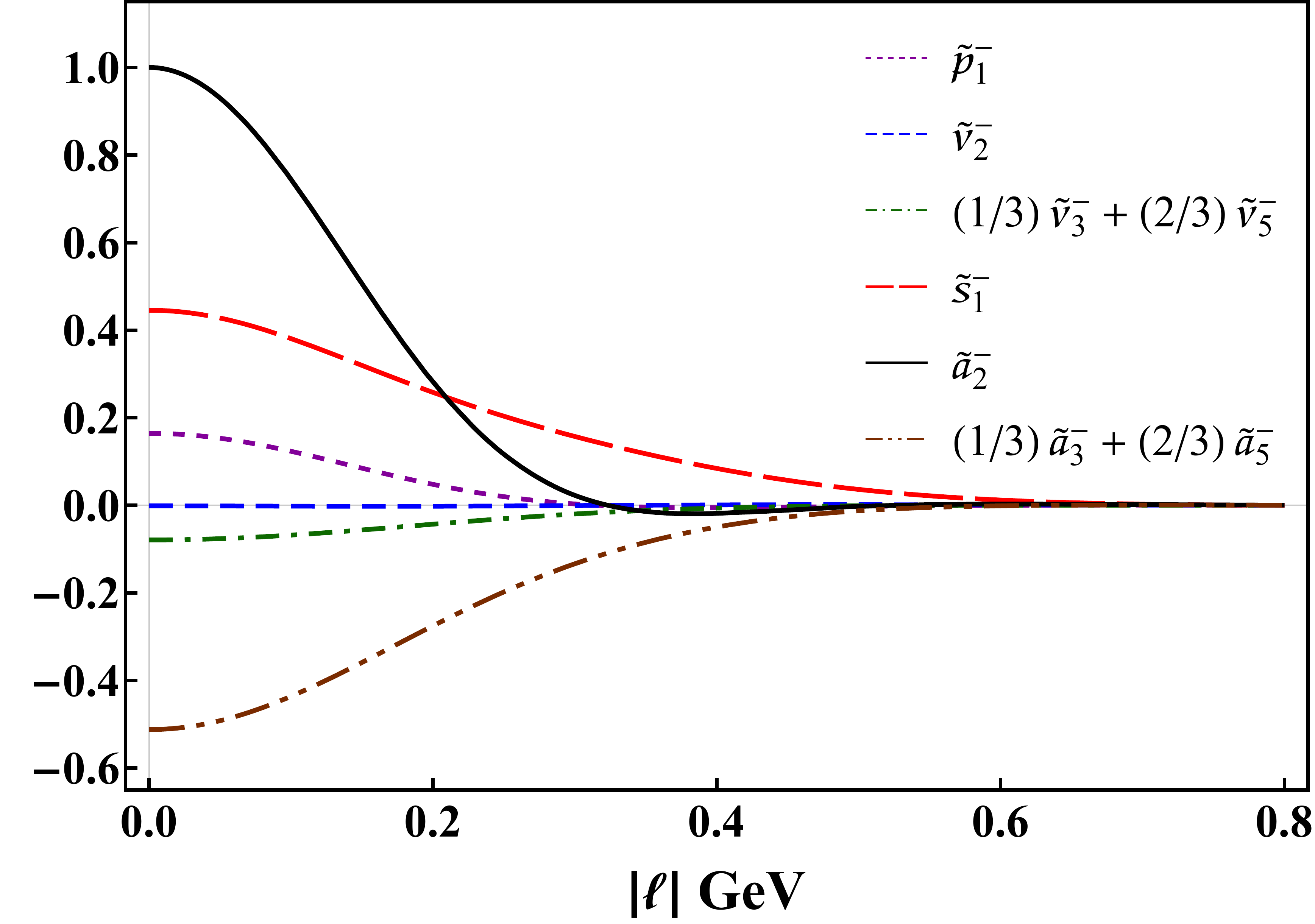}\vspace*{-0ex}
\end{tabular}
\begin{tabular}{lr}
\includegraphics[clip,width=0.4\linewidth]{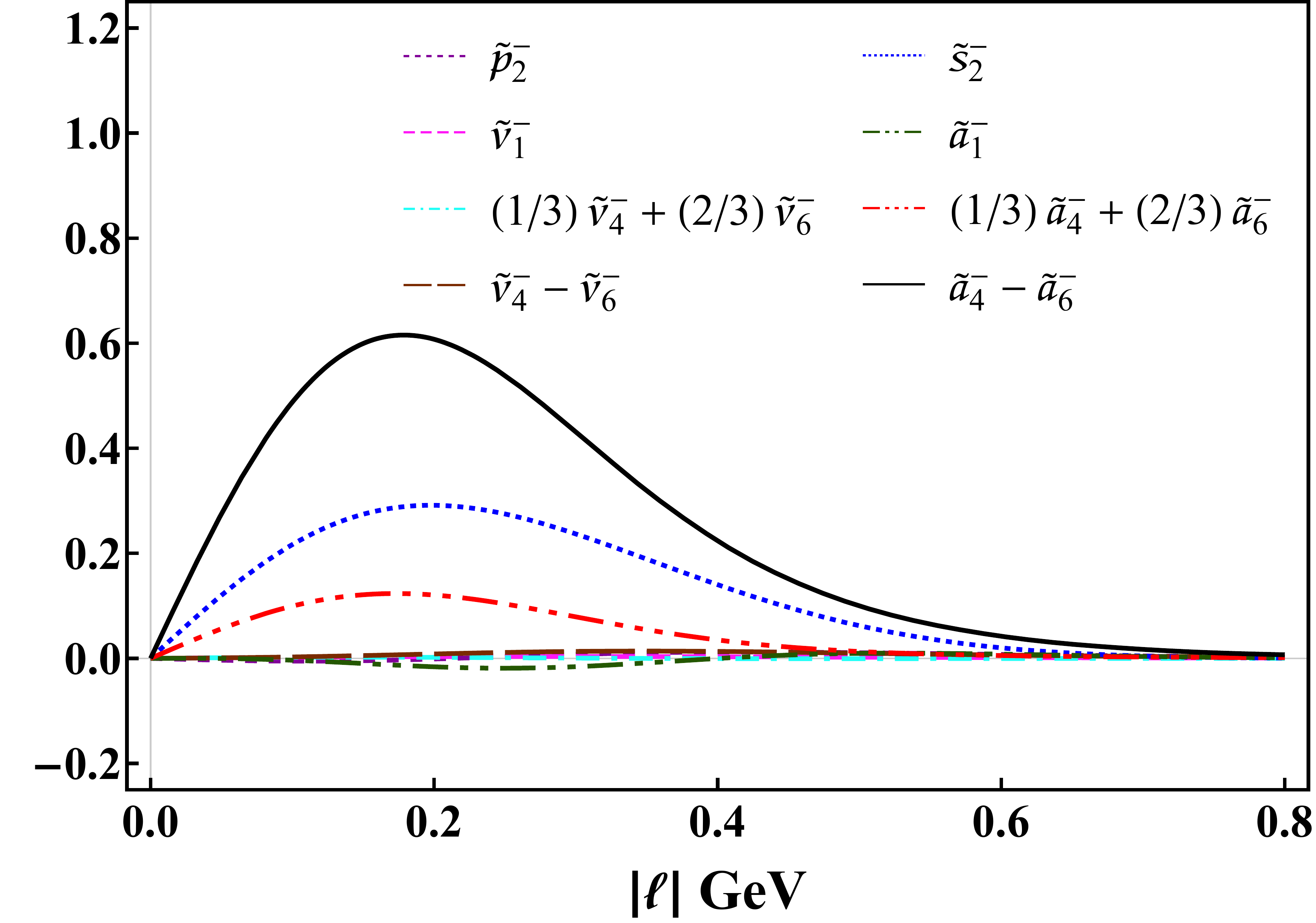}\hspace*{2ex } &
\includegraphics[clip,width=0.4\linewidth]{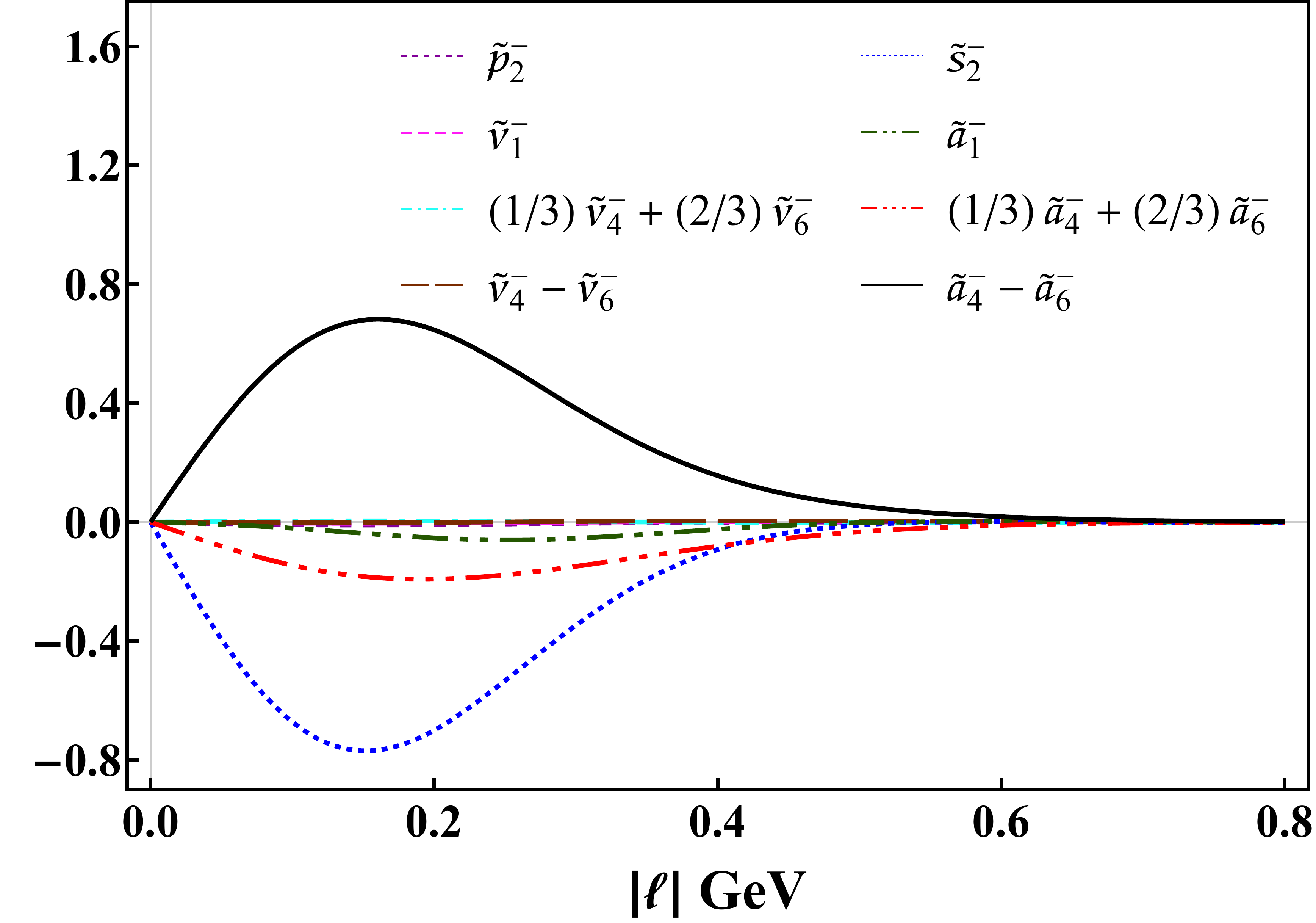}\vspace*{-1ex}
\end{tabular}
\end{center}
\caption{\label{FAPm}
Order-zero Chebyshev projections, Eq.\,\eqref{Wproject}, of the $N(1535)\,1/2^-$ (left) and $N(1650)\,1/2^-$ (right) quark-core Faddeev amplitudes, with $S$-wave in the top row and $P$-wave in the bottom.
For a given baryon, all functions are rescaled by the associated $\ell^2=0$ value of the zeroth moment of $\tilde{\mathpzc a}_2^-$.
}
\end{figure*}

\subsection{Pointwise Structure}
The results described hitherto reveal global (integrated) features of the four lightest $J^P=1/2^\pm$ baryon doublets.  It is also worth exposing aspects of their local structure as it is expressed in the pointwise behaviour of their Faddeev amplitudes.  To this end, we consider the zeroth Chebyshev moment of all $S$- and $P$-wave components in a given baryon's Faddeev amplitude,  \emph{i.e}.\ projections of the form
\begin{equation}
\label{Wproject}
{\mathpzc E}(\ell^2;P^2) = \frac{2}{\pi} \int_{-1}^1 \! du\,\sqrt{1-u^2}\,
{\mathpzc E}(\ell^2,u; P^2)\,,
\end{equation}
where $u=\ell\cdot P/\sqrt{\ell^2 P^2}$.  (The $D$-wave components are uniformly small.)

The order-zero Chebyshev projections of the quark-core Faddeev amplitudes for the nucleon and its positive-parity excitation are plotted in Fig.\,\ref{FAPp}.  Evidently, whilst these projections of the nucleon's Faddeev amplitude are each of a single sign, either positive or negative, those associated with the quark core of the nucleon's first positive-parity excitation are quite different: all $S$-wave components exhibit a single zero at $z_R \approx 0.4\,$GeV$\approx 1/[0.5\,{\rm fm}]$; and four of the $P$-wave projections also possess a zero.
Drawing upon experience with quantum mechanics and with excited-state mesons studied via the Bethe-Salpeter equation \cite{Holl:2004fr, Qin:2011xq, Rojas:2014aka, Li:2016dzv, Li:2016mah}, this pattern of behaviour for the first excited state indicates that it may be interpreted as a radial excitation.  (These observations and conclusions match those in Ref.\,\cite{Segovia:2015hra}.)
%
Notably, too, the relative magnitudes of these Faddeev amplitude projections are consistent with the angular momentum contents indicated by Fig.\,\ref{BarLD}A.

We depict the order-zero Chebyshev projections of the Faddeev amplitudes associated with the $N(1535)\,1/2^-$, $N(1650)\,1/2^-$ quark cores in Fig.\,\ref{FAPm}.  The contrast with the positive-parity states is stark.  In particular, there is no simple pattern of zeros, with all panels containing at least one function that possesses a zero.

Combining the results in Table\,\ref{tableL} and Figs.\,\ref{BarLDN}A, \ref{FAPm}, our analysis indicates that, in their rest frames, the amplitudes associated with these negative-parity states contain roughly equal fractions of even and odd parity diquarks.  Concerning quark-diquark orbital angular momentum, these systems are predominantly $P$-wave in nature, both with strong $ ^2P$ and $ ^4P$ fractions, but possess material $S$-wave components; and the first excited state in this negative parity channel -- $N(1650)\,1/2^-$ -- has little of the appearance of a radial excitation, since most of the functions depicted in the right panels of Fig.\,\ref{FAPm} do not possess a zero.   Similar conclusions may be drawn from the studies in Refs.\,\cite{Eichmann:2016yit, Eichmann:2016hgl}.

These observations provide partial support for the constituent-quark model picture, in which $N(1535)\,1/2^-$, $N(1650)\,1/2^-$ are identified with the $(70,1_1^-)$ supermultiplet, \emph{viz}.\ they are states with a single unit of orbital angular momentum located in one of the two-quark relative coordinates.
With such features, the character of the negative parity baryons produced by our QCD-kindred kernel is markedly different from that generated by a contact interaction \cite{Lu:2017cln}, which suppresses orbital angular momentum and enhances like-parity diquark content.  In particular, although $N(1440)\,1/2^+$ and $N(1650)\,1/2^-$ are naturally identified as parity partners, owing to their appearance as the second states in the $1/2^\pm$ channels, respectively, they are remarkably different in structure.
 %

The structural dissimilarity just described suggests that the mere observation of a collection of (nearly) degenerate parity-partners above some mass-scale is insufficient to claim the restoration of chiral symmetry at and above that scale in the hadron spectrum.  Although similar in mass, the structure of opposite-parity partner states might nevertheless be very different, in which case DCSB would still be playing a decisive role and numerous other measurable properties would remain as signals to distinguish between the partners.  Such an outcome is particularly likely if there is a tight link between DCSB and dynamical quark confinement in QCD \cite{Roberts:2016vyn}; and we have seen this in preliminary investigations of higher excitations in the $(1/2,1/2^\pm)$ channels.

Complementing Refs.\,\cite{Burkert:2004sk, Aznauryan:2012ba, Anikin:2015ita, Proceedings:2016jkcA, Proceedings:2016jkcB, Aznauryan:2017nkz}, our analysis indicates that the pointwise behaviour of nucleon-to-resonance electroproduction form factors, \emph{e.g}.\ $N \to N(1535)\,1/2^-$, $N\to N(1650)\,1/2^-$, on $Q^2 \gtrsim 2\,$GeV$^2$ should serve well in discriminating between otherwise viable pictures of baryon and resonance structure, as has already been found with the $N(1440)\,1/2^+$ \cite{Burkert:2017djo}.
In this connection, experimental results for the $N \to N(1535)\,1/2^-$ electromagnetic transition amplitudes on $Q^2 \gtrsim 2\,$GeV$^2$ are already available \cite{Aznauryan:2009mx} and they are expected soon from data on $N \to N(1650)\,1/2^-$ \cite{Isupov:2017lnd}.



\section{Summary}
\label{epilogue}
Using a Faddeev kernel that is known to support a uniformly good description of the observed properties of the nucleon, $\Delta$-baryon and the Roper resonance, we performed a comparative study of the four lightest $(I=1/2,J^P = 1/2^\pm)$ baryon isospin-doublets in order to both elucidate their structural similarities and differences, and draw whatever relationships might exist with quark model descriptions of these systems.

A basic prediction of such Faddeev equation studies is the presence of strong nonpointlike, fully-interacting quark-quark (diquark) correlations within all baryons.  In keeping with earlier studies, we found that a complete description of the two lightest $(I=1/2,J^P=1/2^+)$ doublets is obtained by retaining only isoscalar-scalar $(I=0,J^P=0^+)$ and isovector-pseudovector correlations, \emph{i.e}.\ even allowing for the possibility of isoscalar-pseudoscalar and -vector correlations, strong interaction dynamics in the $1/2^+$ baryon channels ensure that pseudoscalar and vector diquarks play a negligible role in forming the bound states.  Consequently, the Faddeev amplitudes which describe the dressed-quark cores of the two lightest $(I=1/2,J^P=1/2^+)$ doublets are dominated by scalar and pseudovector diquarks; the associated rest-frame Faddeev wave functions are primarily $S$-wave in nature; and the first excited state in this $1/2^+$ channel has very much the appearance of a radial excitation of the ground state.

In connection with the two lightest $(I=1/2,J^P=1/2^-)$ doublets, one might imagine that the situation is reversed, \emph{viz}.\ that isoscalar-pseudoscalar and -vector correlations are the dominant diquark constituents.  However, this is not the case.  In these systems, too, scalar and pseudovector diquarks play a material role.
Indeed, a good approximation to their masses is obtained by retaining solely pseudovector correlations; in their rest frames, the Faddeev amplitudes describing the dressed-quark cores of these negative-parity states contain roughly equal fractions of even and odd parity diquarks; the associated wave functions of these negative-parity systems are predominantly $P$-wave in nature, both with strong $ ^2P$ and $ ^4P$ fractions, but possess measurable $S$-wave components; and, interestingly, the first excited state in this negative parity channel has little of the appearance of a radial excitation: instead, it is distinguished from the ground-state by its angular momentum structure.

There are some similarities here with quark model descriptions of these systems, so long as rest-frame orbital angular momentum is identified with that existing between dressed-quarks and -diquarks, which are the correct strong-interaction quasiparticle degrees-of-freedom at the hadronic scale and on a material domain extending beyond.  On the other hand, it is important to stress that in our quantum field theory analysis the negative parity states are not purely angular-momentum excitations of the $(1/2,1/2^+)$ ground-state.  Their Faddeev wave functions contain both $P$- and $S$-wave components, and also express some features of radial excitations.

To test these pictures, we introduced a parameter, $g_{\rm DB}$: reducing its value from unity worked to suppress the role of opposite-parity diquarks in the Faddeev kernel of a given $1/2^P$ baryon.  Its value had no impact on the $P=+$ systems; and in $P=-$ systems, a value of $g_{\rm DB}<0.1$ was needed in order to enforce dominance of pseudoscalar and vector diquarks.  It appears, therefore, that the four lightest $(I=1/2,J^P=1/2^\pm)$ doublets are indeed primarily constituted from even-parity diquarks and hence the findings described above are robust.

At this point it is worth reiterating that the interpolating fields for positive and negative parity states may simply be related by chiral rotation of the quark spinors used in their construction.  Hence, any differences between the bound-state equations and their solutions in these channels are generated by chiral symmetry breaking, which is overwhelmingly dynamical in the light-quark sector.  In the present context, this entails that the following pairs are parity partners: $N(940)\,1/2^+$-$N(1535)\,1/2^-$, $N(1440)\,1/2^+$-$N(1650)\,1/2^-$.  It is common to ascribe the mass-splitting between such parity partners to dynamical chiral symmetry breaking (DCSB); but our analysis reveals very material differences between their internal structure, too, and those differences must also be attributable to DCSB because the channels are identical when chiral symmetry is restored.  Since a tight connection very probably exists between DCSB and confinement in the Standard Model, then experiments which can test the contrasts we have drawn between the internal structure of the four lightest $(I=1/2,J^P = 1/2^\pm)$ doublets will serve a very valuable purpose.  In this connection, resonance electroproduction experiments on $Q^2 \gtrsim 2\,$GeV$^2$ provide one clear example.

\acknowledgments
We are grateful for constructive comments from R.~Gothe, G.~Krein, Y.~Lu, C.~Mezrag, V.~Mokeev and S.-X. Qin, and for the hospitality and support of RWTH Aachen University, III.\,Physikalisches Institut B, Aachen, Germany.
Work supported by:
Funda\c{c}\~ao de Amparo \`a Pesquisa do Estado de S\~ao Paulo - FAPESP Grant No. 2015/21550-4;
U.S.\ Department of Energy, Office of Science, Office of Nuclear Physics, under contract no.~DE-AC02-06CH11357;
Chinese Ministry of Education, under the \emph{International Distinguished Professor} programme;
European Union's Horizon 2020 research and innovation programme under the Marie Sk\l{}odowska-Curie Grant Agreement No.\ 665919;
Spanish MINECO's Juan de la Cierva-Incorporaci\'on programme, Grant Agreement No. IJCI-2016-30028;
Spanish Ministerio de Econom\'ia, Industria y Competitividad under Contract Nos.\ FPA2014-55613-P and SEV-2016-0588;
Forschungszentrum J\"ulich GmbH;
and National Natural Science Foundation of China, Contract No.\,11275180.


\appendix
\setcounter{equation}{0}
\setcounter{figure}{0}
\setcounter{table}{0}
\renewcommand{\theequation}{\Alph{section}.\arabic{equation}}
\renewcommand{\thetable}{\Alph{section}.\arabic{table}}
\renewcommand{\thefigure}{\Alph{section}.\arabic{figure}}

\section{Dressed quark propagator}
\label{appendixKFE}
The dressed-quark propagator can be written:
\begin{subequations}
\begin{align}
S(p) & =  -i \gamma\cdot p\, \sigma_V(p^2) + \sigma_S(p^2) \\
& = 1/[i\gamma\cdot p\, A(p^2) + B(p^2)]\,.
\label{SpAB}
\end{align}
\end{subequations}
It is known that for light-quarks the wave function renormalisation and dressed-quark mass:
\begin{equation}
\label{ZMdef}
Z(p^2)=1/A(p^2)\,,\;M(p^2)=B(p^2)/A(p^2)\,,
\end{equation}
respectively, receive strong momentum-dependent corrections at infrared momenta \cite{Lane:1974he, Politzer:1976tv, Zhang:2004gv, Bhagwat:2004kj, Bhagwat:2006tu, Binosi:2016wcx}: $Z(p^2)$ is suppressed and $M(p^2)$ enhanced.  These features are an expression of DCSB and, plausibly, of confinement \cite{Horn:2016rip}; and their impact on hadron phenomena has long been emphasised \cite{Roberts:1994hh}.

\begin{figure}[!t]
\centerline{%
\includegraphics[clip, width=0.45\textwidth]{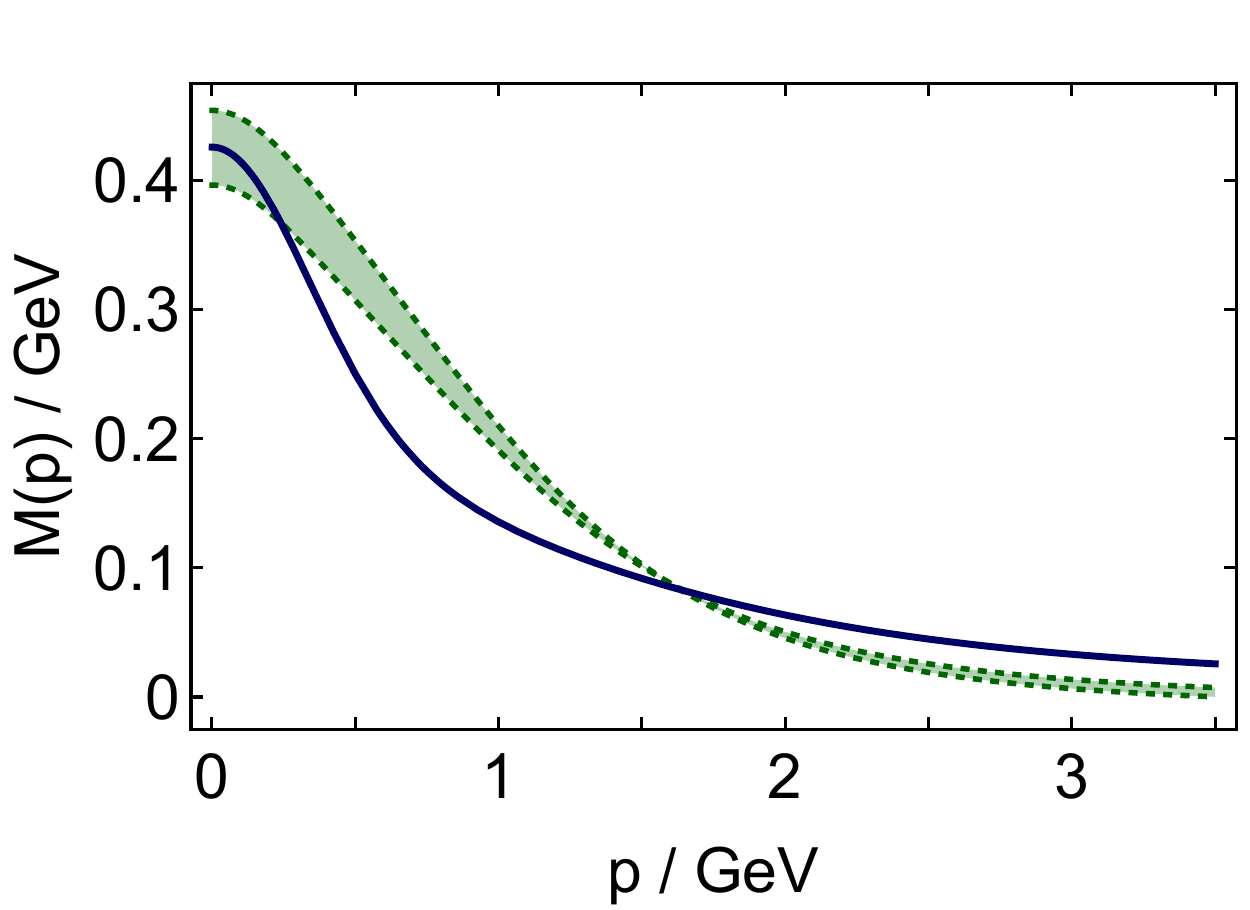}}
\caption{\label{MassPlot}
Solid curve (blue) -- quark mass function generated by the parametrisation of the dressed-quark propagator specified by Eqs.\,\eqref{EqSSSV}--\eqref{tableA}; and band (green) -- exemplary range of numerical results obtained by solving the gap equation with the modern DCSB-improved kernels described and used in Refs.\,\cite{Chang:2010hb, Chang:2011ei, Chang:2013pq, Chang:2013epa}.}
\end{figure}

Numerical solutions of the quark gap equation are now readily obtained.  However, the utility of an algebraic form for $S(p)$ when calculations require the evaluation of numerous multidimensional integrals is self-evident.  An efficacious parametrisation of $S(p)$, which exhibits the features described above, has been used extensively in hadron studies \cite{Roberts:2007jh}.  It is expressed via
{\allowdisplaybreaks
\begin{subequations}
\label{EqSSSV}
\begin{align}
\bar\sigma_S(x) & =  2\,\bar m \,{\cal F}(2 (x+\bar m^2)) \nonumber \\
& \quad + {\cal F}(b_1 x) \,{\cal F}(b_3 x) \,
\left[b_0 + b_2 {\cal F}(\epsilon x)\right]\,,\label{ssm} \\
\label{svm} \bar\sigma_V(x) & =  \frac{1}{x+\bar m^2}\, \left[ 1 - {\cal F}(2
(x+\bar m^2))\right]\,,
\end{align}
\end{subequations}}
\hspace*{-0.5\parindent}with $x=p^2/\lambda^2$, $\bar m$ = $m/\lambda$,
\begin{equation}
\label{defcalF}
{\cal F}(x)= \frac{1-\mbox{\rm e}^{-x}}{x}  \,,
\end{equation}
$\bar\sigma_S(x) = \lambda\,\sigma_S(p^2)$ and $\bar\sigma_V(x) =
\lambda^2\,\sigma_V(p^2)$.
The mass-scale, $\lambda=0.566\,$GeV, and
parameter values
\begin{equation}
\label{tableA}
\begin{array}{ccccc}
   \bar m& b_0 & b_1 & b_2 & b_3 \\\hline
   0.00897 & 0.131 & 2.90 & 0.603 & 0.185
\end{array}\;,
\end{equation}
associated with Eqs.\,\eqref{EqSSSV} were fixed in a least-squares fit to light-meson observables \cite{Burden:1995ve, Hecht:2000xa}.  ($\epsilon=10^{-4}$ in Eq.\ (\ref{ssm}) acts only to decouple the large- and intermediate-$p^2$ domains.)

The dimensionless $u=d$ current-quark mass in Eq.\,(\ref{tableA}) corresponds to
$m=5.08\,{\rm MeV}$ 
and the parametrisation yields the following Euclidean constituent-quark mass, defined as the solution of $p^2=M^2(p^2)$:
$M_{u,d}^E = 0.33\,{\rm GeV}$.
The ratio $M^E/m = 65$ is one expression of DCSB in the parametrisation of $S(p)$.  It emphasises the dramatic enhancement of the dressed-quark mass function at infrared momenta.

The dressed-quark mass function generated by this parametrisation is depicted in Fig.\,\ref{MassPlot}, wherein it is compared with that computed using the DCSB-improved gap equation kernel described in Refs.\,\cite{Chang:2010hb, Chang:2011ei} and used subsequently to predict the pion parton distribution amplitudes form factors \cite{Chang:2013pq, Chang:2013epa}.  Evidently, although simple and introduced long beforehand, the parametrisation is a sound representation of contemporary numerical results.  (We note that the numerical solutions depicted in Fig.\,\ref{MassPlot} were obtained in the chiral limit, which explains why the (green) band in falls below the parametrisation at larger $p$.)

As with the diquark propagators in Eq.\,\eqref{Eqqqprop}, the expressions in Eq.\,\eqref{EqSSSV} ensure confinement of the dressed quarks via the violation of reflection positivity (see, \emph{e.g}.\ Ref.\,\cite{Horn:2016rip}, Sec.\,3).


\end{document}